\documentclass[10pt,conference]{IEEEtran}
\usepackage{cite}
\usepackage{amsmath,amssymb,amsfonts}
\usepackage{algorithmic}
\usepackage{graphicx}
\usepackage{textcomp}
\usepackage{xcolor}
\usepackage[hyphens]{url}
\usepackage{fancyhdr}
\usepackage{hyperref}
\usepackage{xspace}
\usepackage{tikz}
\usepackage{multicol}

\newcommand{\hpcayear}{2026}

\newcommand*\circled[2]{\tikz[baseline=(char.base)]{
            \node[shape=circle,fill=black,inner sep=1pt] (char) {\textcolor{#1}{{\footnotesize #2}}};}}

\ifx\figurename\undefined \def\figurename{Figure}\fi
\renewcommand{\figurename}{Fig.}
\renewcommand{\paragraph}[1]{\textbf{#1} }
\newcommand{\subparagraph}[1]{\underline{\textit{#1}} }

\newcommand{\Sect}[1]{Sec.~\ref{#1}}
\newcommand{\Fig}[1]{Fig.~\ref{#1}}

\newcommand{\Eqn}[1]{Eqn.~\ref{#1}}

\newcommand{\Review}[1]{}

\def\cF{{\mathcal{F}}}

\def\cP{{\mathcal{P}}}

\newcommand{\mode}[1]{\underline{\textsc{#1}}\xspace}

\newcommand{\proj}{\textsc{Splatonic}\xspace}

\graphicspath{{figs/}}

\newcommand{\hpcasubmissionnumber}{176}
\title{\proj: Architectural Support for 3D Gaussian Splatting SLAM via Sparse Processing}
\def\hpcacameraready{} % Uncomment to build camera-ready version

\author{
  \ifdefined\hpcacameraready
    \IEEEauthorblockN{Xiaotong Huang\textsuperscript{1,\dag}, He Zhu\textsuperscript{1,\dag}, Tianrui Ma\textsuperscript{3}, Yuxiang Xiong\textsuperscript{1}, Fangxin Liu\textsuperscript{1} \\ Zhezhi He\textsuperscript{1},
    Yiming Gan\textsuperscript{3}, Zihan Liu\textsuperscript{1, 2, *}, Jingwen Leng\textsuperscript{1, 2}, Yu Feng\textsuperscript{1, 2, *}, Minyi Guo\textsuperscript{1, 2}}
    
    \IEEEauthorblockA{
        \textsuperscript{1}Shanghai Jiao Tong University, \textsuperscript{2}Shanghai Qi Zhi Institute \\
        \textsuperscript{3}Institute of Computing Technology, Chinese Academy of Science}
    \IEEEauthorblockA{
        \{hxt0512, zhcon16, xiongyuxiang, liufangxin, zhezhi.he, altair.liu, leng-jw, y-feng, guo-my\}@sjtu.edu.cn \\
        \{matianrui, ganyiming\}@ict.ac.cn
      }
    \IEEEauthorblockA{
        $\dagger$Equal contribution, *Corresponding authors 
    }
    \vspace{2mm}
    Project site: \url{https://stonesix16.github.io/splatonic/}
  \else
    \IEEEauthorblockN{\normalsize{HPCA \hpcayear{} Submission
      \textbf{\#\hpcasubmissionnumber{}}} \\
      \IEEEauthorblockA{
        Confidential Draft \\
        Do NOT Distribute!!
      }
    }
  \fi 
}

\fancypagestyle{camerareadyfirstpage}{%
  \fancyhead{}
  
  \fancyhead[C]{
    \ifdefined\aeopen
    \parbox[][12mm][t]{13.5cm}{\hpcayear{} IEEE International Symposium on High-Performance Computer Architecture (HPCA)}    
    \else
      \ifdefined\aereviewed
      \parbox[][12mm][t]{13.5cm}{\hpcayear{} IEEE International Symposium on High-Performance Computer Architecture (HPCA)}
      \else
      \ifdefined\aereproduced
      \parbox[][12mm][t]{13.5cm}{\hpcayear{} IEEE International Symposium on High-Performance Computer Architecture (HPCA)}
      \else
      \parbox[][0mm][t]{13.5cm}{\hpcayear{} IEEE International Symposium on High-Performance Computer Architecture (HPCA)}
    \fi 
    \fi 
    \fi 
    \ifdefined\aeopen 
      \includegraphics[width=12mm,height=12mm]{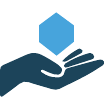}
    \fi 
    \ifdefined\aereviewed
      \includegraphics[width=12mm,height=12mm]{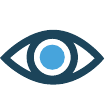}
    \fi 
    \ifdefined\aereproduced
      \includegraphics[width=12mm,height=12mm]{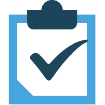}
    \fi
  }
  \fancyfoot[C]{}
}
\ifdefined\hpcacameraready 
\else
\fi

\ifdefined\eaopen

\fi

\begin{document}

\maketitle
%%%%%%%%%%%%%%%%%%%%%%%%%%%%%%%%%%%%%%%%
%%%%%%%% -- PAPER CONTENT STARTS -- %%%%%%%%%

\begin{abstract}

3D Gaussian splatting (3DGS) has emerged as a promising direction for SLAM due to its high-fidelity reconstruction and rapid convergence. 
However, 3DGS-SLAM algorithms remain impractical for mobile platforms due to their high computational cost, especially for their tracking process. 

This work introduces \proj, a sparse and efficient real-time 3DGS-SLAM algorithm-hardware co-design for resource-constrained devices. 
Inspired by classical SLAMs, we propose an adaptive sparse pixel sampling algorithm that reduces the number of rendered pixels by up to 256$\times$ while retaining accuracy. 
To unlock this performance potential on mobile GPUs, we design a novel pixel-based rendering pipeline that improves hardware utilization via Gaussian-parallel rendering and preemptive $\alpha$-checking.
Together, these optimizations yield up to 121.7$\times$ speedup on the bottleneck stages and 14.6$\times$ end-to-end speedup on off-the-shelf GPUs. 
To further address new bottlenecks introduced by our rendering pipeline, we propose a pipelined architecture that simplifies the overall design while addressing newly emerged bottlenecks in projection and aggregation. 
Evaluated across four 3DGS-SLAM algorithms, \proj achieves up to 274.9$\times$ speedup and 4738.5$\times$ energy savings over mobile GPUs and up to 25.2$\times$ speedup and 241.1$\times$ energy savings over state-of-the-art accelerators, all with comparable accuracy.

\end{abstract}


\section{Introduction}
\label{sec:intro}

% Simultaneous localization and mapping (SLAM) is a key component to intelligent automation across various applications, including embodied AI~\cite{duan2022survey}, autonomous driving~\cite{bresson2017simultaneous}, augmented/virtual reality (AR/VR)~\cite{sheng2024review}, smart factory~\cite{wu2023rf}, and many more~\cite{kazerouni2022survey}.
Simultaneous localization and mapping (SLAM) is a key component to intelligent automation across various applications~\cite{duan2022survey, bresson2017simultaneous, sheng2024review, wu2023rf, kazerouni2022survey}.
Among SLAM algorithms, 3D Gaussian splatting (3DGS)-based algorithms~\cite{keetha2024splatam, yan2024gs, yugay2023gaussian, matsuki2024gaussian, pham2024flashslam} have recently emerged as a promising direction due to their superior reconstruction fidelity and fast convergence~\cite{tosi2024nerfs, zhang2024hi, bai2024rp, wang2024new}, over their alternatives.
For example, iterative closest point (ICP)-based methods~\cite{zhang2014loam, newcombe2011kinectfusion, hess2016real} often struggle in low-texture and poor lighting environments.
Neural radiance field (NeRF)-based methods~\cite{zhu2022nice, johari2023eslam, sandstrom2023point, gong2025hs} suffer from high compute costs and slow rendering.

Nevertheless, 3DGS-SLAM algorithms remain constrained in mobile applications due to their substantial computation overheads.
The state-of-the-art 3DGS-SLAM algorithms often fail to achieve real-time on off-the-shelf mobile SoCs~\cite{orin_nx, orinsoc, xaviersoc}.
For instance, the average frame rate of a 3DGS-SLAM algorithm, SplaTAM~\cite{keetha2024splatam}, is only 0.1 Hz on a mobile Ampere GPU~\cite{orinsoc}, far from real-time (10-30 Hz).
This gap between the computation demands and the hardware capability motivates the need for an acceleration solution on mobile devices.

In general, 3DGS-SLAM algorithms consist of two concurrent processes: \textit{tracking} and \textit{mapping}. 
Tracking estimates the camera pose of each frame based on the reconstructed scene, while mapping reconstructs unseen regions of the scene by generating new 3D Gaussian primitives.
These two processes often operate at different frequencies: tracking runs in a per-frame manner to ensure the pose accuracy; and mapping is invoked less frequently, usually every 4-8 frames.
Our experiment shows that the amortized per-frame latency of mapping is only one-quarter that of tracking (\Fig{fig:latency}). 
Thus, our work primarily focuses on accelerating the tracking process.

\paragraph{Idea.}
Although tracking and mapping serve different purposes, both share the same differentiable rendering pipeline and rely on training to obtain accurate results.
Among all stages in this pipeline, \textit{rasterization} is the primary bottleneck in both forward and backward training passes (see \Fig{fig:exec_time}), accounting for 94.7\% of the execution time, due to the need to iterate over every pixel of a frame in rasterization. 
Inspired by classical SLAM approaches~\cite{zhang2014loam, newcombe2011kinectfusion, hess2016real}, which accelerate localization by detecting key features, we explore whether the similar sparsity-based principle can apply to 3DGS-SLAMs. 
Since the workload of rasterization is proportional to the number of processed pixels, processing fewer pixels could dramatically reduce the overall computation cost~\cite{kerbl20233d}. 

Thus, we propose an adaptive sparse sampling algorithm in \Sect{sec:algo:sample} that can select important pixels at runtime to reduce the overall computation cost. 
Based on the characteristics of tracking and mapping, we tailor sampling strategies for each process.
In tracking, we find that a simple yet effective approach, uniform random sampling, is sufficient to preserve pose estimation accuracy, while reducing the number of processed pixels by 256$\times$.
In mapping, on the other hand, our sampling algorithm prioritizes pixels in unseen regions or areas with rich textures to preserve reconstruction quality.
Overall, we show that our sampling algorithm achieves the best accuracy compared to existing sampling strategies.

\paragraph{Pipeline.} 
However, directly applying our sparse sampling algorithm to the existing 3DGS pipeline achieves merely a 4.2$\times$ speedup on rasterization, far below the compute savings (256$\times$).
The fundamental reason is that current 3DGS pipelines~\cite{kerbl20233d, fang2024mini, wang2024adr, huang2025seele} all adopt a \textit{tile-based} rendering, which exploits the data sharing across pixels in rasterization to amortize the compute cost of its earlier stages, i.e., projection and sorting.
However, this rendering paradigm is inherently \textit{ill-suited} to sparse pixel rendering, as sparsely distributed pixels offer little opportunity for data sharing and result in low hardware utilization on both GPUs and dedicated accelerators.

Thus, we design a \textit{pixel-based} rendering pipeline in \Sect{sec:algo:pipeline}. 
Compared to tile-based rendering, our pipeline has two key advantages.
First, we explicitly perform pixel-level projection that eliminates the unnecessary computation of subsequent stages that would otherwise result from the data sharing of tile-based rendering.
Second, we propose a Gaussian-parallel rasterization, where multiple processing elements (PEs) co-render a single pixel, rather than assigning one pixel per PE.
Our Gaussian-parallel rasterization largely improves the PE parallelism.
% and tames warp divergence.
In addition to our pixel-based rendering, we further propose \textit{preemptive $\alpha$-checking}, a technique to eliminate the warp divergence across PEs and avoid the unnecessary computations after the projection stage.
% Conventional 3DGS pipelines apply $\alpha$-checking in rasterization to avoid the computation of unnecessary Gaussians from data sharing across pixels. 
% However, $\alpha$-checking accounts for over 50\% of rasterization time and is the primary source of warp divergence (\Sect{sec:motiv:ch}).
% Our Gaussian-parallel rasterization no longer exploits this data sharing; thus, we preemptively move $\alpha$-checking to projection.
% This way, there is no warp divergence, and we further remove the unnecessary Gaussians in subsequent sorting and rasterization. 
With all optimizations together, we boost the rasterization performance up to 121.7$\times$ on an off-the-shelf GPU in \Sect{sec:eval:gpu}. 

\paragraph{Architectural Support.} 
With rasterization no longer the bottleneck, we find that the main bottlenecks shift to the projection in the forward pass and the aggregation in the backward pass (\Fig{fig:algo_overview}).
To address new bottlenecks, we propose a clean-slate pipelined architecture in \Sect{sec:arch}.
Specifically, we introduce a lightweight rasterization engine that simplifies the rendering logic of rasterization in both the forward and backward passes. 
Furthermore, we augment a projection unit to mitigate the increased computational overhead by preemptive $\alpha$-checking. 
Finally, we design a specialized aggregation unit to address the frequent pipeline stalls due to aggregation.

\paragraph{Result.} We evaluate \proj on four popular 3DGS-SLAM algorithms. 
With our sampling algorithm and pixel-based rendering, \proj achieves 14.6$\times$ end-to-end speedup and 86.1\% energy reduction on the mobile Ampere GPU~\cite{orinsoc} with comparable accuracy against the baseline algorithms.
With hardware support, \proj achieves up to 274.9$\times$ speedup and 4738.5$\times$ energy savings against the GPU baseline, while achieving up to 25.2$\times$ speedup and 241.1$\times$ energy savings against the prior accelerators~\cite{he2025gsarch, wu2024gauspu}.
Even with the same sparse sampling algorithm, \mbox{\proj} still can achieve up to 12.7{$\times$} speedup and 200.8{$\times$} energy savings against prior accelerators~\mbox{\cite{he2025gsarch, wu2024gauspu}}.

The contributions of this paper are as follows:
\begin{itemize}
    \item We propose a sparse pixel sampling algorithm for 3DGS-SLAMs that achieves up to 256$\times$ pixel reduction in tracking with an even better task accuracy.
    \item We introduce a pixel-based rendering pipeline that improves the parallelism in sparse pixel processing and achieves up to 121.7$\times$ speedup on the bottleneck stages.
    \item We co-design a pipelined accelerator architecture, which further addresses the remaining bottlenecks in pixel-based rendering and further improves performance.
\end{itemize}

\section{Background}
\label{sec:bg}

\subsection{3DGS-based SLAM}
\label{sec:bg:slam}

\begin{figure}[t]
    \centering
    \includegraphics[width=\columnwidth]{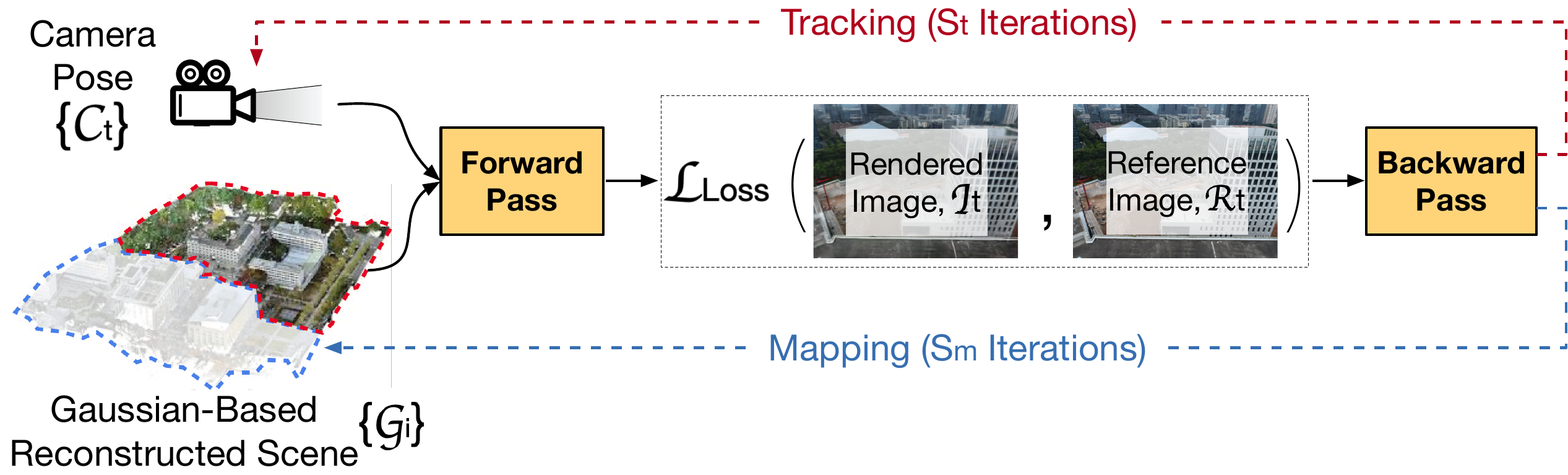}
    \caption{Overview of 3DGS-SLAM process. Tracking and mapping share the same optimization pipeline with different optimization targets. Tracking optimizes camera poses $\{C_t\}$ while mapping reconstructs the scene $\{G_i\}$.}
    \label{fig:slam_overview}
\end{figure}

\paragraph{Overview.}
The overall goal of SLAM is to reconstruct the scene of an unknown environment while simultaneously estimating the agent's trajectory within that scene.
3DGS-SLAM algorithms~\cite{keetha2024splatam, yugay2023gaussian, matsuki2024gaussian, pham2024flashslam} achieve this by leveraging 3DGS rendering pipelines, as illustrated in \Fig{fig:slam_overview}.
The overall process can be divided into two concurrent processes: \textit{tracking} and \textit{mapping}.
These two processes jointly optimize two sets of trainable parameters: the camera trajectory, $\{C_t\}$, and a set of Gaussian points, $\{G_i\}$, that represent the reconstructed scene.
We next describe these two processes separately.

\paragraph{Tracking.}
The goal of tracking is to estimate each camera pose, $C_t$, along the trajectory.
During tracking, we assume that the current reconstructed scene is valid, as shown by the red dashed block in \Fig{fig:slam_overview}.
Under this assumption, we fix the trainable parameters of the Gaussian representation, $\{G_i\}$, and render an image, $I_t$, at the current estimated camera pose, $C_t$, in the forward pass.
By calculating the loss between the rendered image, $I_t$, and the reference image, $R_t$, the backward pass of the 3DGS pipeline then back-propagates the loss to the unfixed parameters, i.e., $C_t$, in a self-supervised manner.
By $S_t$ iterations, the estimated camera pose $C_t$ progressively converges to a value that is close to the true pose.
% The value of $S_t$ is typically adaptive, ranging from \fixme{20} to \fixme{40}.

\begin{figure}[t]
    \centering
    \includegraphics[width=\columnwidth]{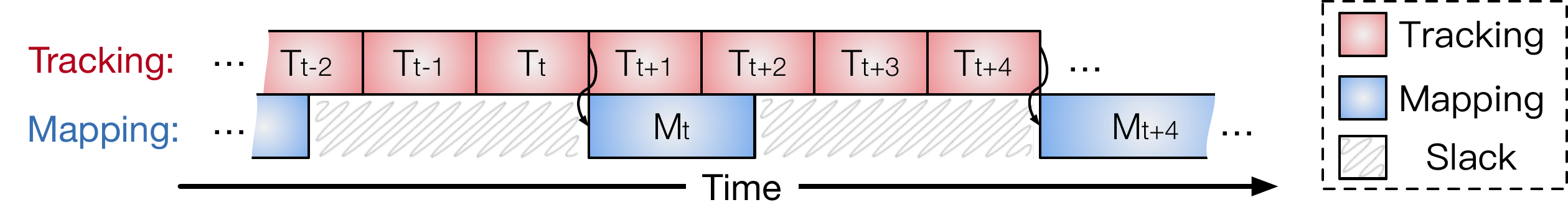}
    \caption{The timing diagram of 3DGS-SLAM process. Tracking often runs more frequently compared to mapping.
    Mapping, $M_t$, at the same time, t, needs to be executed after tracking, $T_t$, due to the dependency.}
    \label{fig:slam_timing}
\end{figure}

\begin{figure*}
    \centering
    \includegraphics[width=\textwidth]{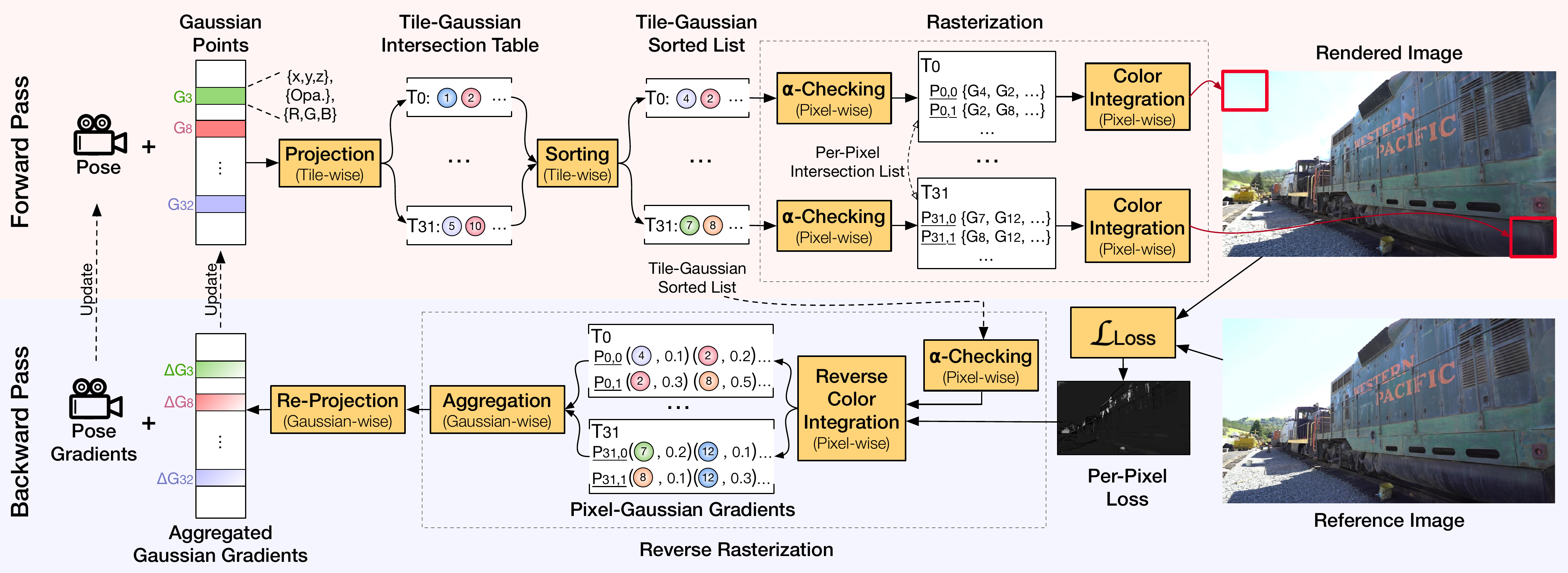}
    \caption{The overview of 3DGS forward and backward passes. 
    The forward pass consists of three stages: \textit{projection}, \textit{sorting}, and \textit{rasterization}. 
    Both projection and sorting are performed at tile granularity to amortize the computational cost across pixels, while rasterization must be performed at the pixel level to render individual pixels correctly.
    Because different pixels within a tile need to integrate different subsets of Gaussians.
    The backward pass mainly comprises two stages: \textit{reverse rasterization} and \textit{re-projection}. 
    Reverse rasterization computes the partial gradients of all pixel-Gaussian pairs and aggregates them to the corresponding Gaussians.
    Re-projection then transforms the accumulated gradients from the camera coordinate system to the world coordinate system.
    }
    \label{fig:algo_overview}
\end{figure*}

\paragraph{Mapping.}
The purpose of mapping, on the other hand, is to reconstruct previously unreconstructed regions of the scene (e.g., the blue dashed blocks in \Fig{fig:slam_overview}) based on the current observations. 
At time $t$, we select $w$ recent camera poses and fix their pose parameters. 
We then fine-tune the 3D Gaussian representations using these poses and their corresponding images. 
This fine-tuning follows the same training process as tracking, except that mapping updates only Gaussian parameters $\{G_i\}$ rather than $\{C_t\}$. 
Over $S_m$ iterations, the Gaussian representation is progressively refined by inserting new Gaussian points until convergence. 
Note that, mapping is typically performed less frequently than tracking.

\paragraph{Order.}
\Fig{fig:slam_timing} shows the timing diagram of a 3DGS-SLAM process.
As shown, tracking and mapping are executed concurrently but at different frequencies.
Tracking (in red) runs continuously at each frame to estimate the camera pose.
On the other hand, mapping (in blue) is invoked less frequently.
However, mapping, $M_t$, at time $t$ must be executed after $T_t$ due to the dependency between tracking and mapping.

\subsection{3DGS Training Pipeline.}
\label{sec:bg:3dgs}

The core of the 3DGS-SLAM algorithm is the 3DGS training process, which maintains two sets of tunable parameters: the camera poses that capture the overall trajectory, $\{C_t\}$, and the 3D Gaussian points, $\{G_i\}$, that model the reconstructed scenes.
Each Gaussian has a set of attributes that capture the geometric and textural properties of the scene.
As shown in \Fig{fig:algo_overview}, the training process can be classified into two passes: the \textit{forward pass} and the \textit{backward pass}.

% The forward pass renders an image given the current pose estimate, while the backward pass computes the loss between the rendered image and the reference image, and backpropagates this loss to update the tunable parameters.
% We next describe these two passes.

\paragraph{Forward Pass.}
The overall forward pass consists of three stages: \textit{projection}, \textit{sorting}, and \textit{rasterization}.
% Today's 3DGS forward pipeline all adopt the tile-based rendering.

\subparagraph{Projection.}
The purpose of projection is to filter out Gaussians that lie outside the current view frustum.
To amortize computational overhead, current pipelines perform this filtering at the tile level, rather than pixel-by-pixel, to identify the intersection between Gaussians and rendering tiles.
The results are then written into a tile-Gaussian intersection table.

\subparagraph{Sorting.}
Once each tile obtains its intersected Gaussians, sorting determines the rendering order of those Gaussians within individual tiles.
This stage ensures that all Gaussians are rendered in a correct order, from the closest to the farthest. 

\subparagraph{Rasterization.}
The sorted Gaussians in the tile-Gaussian sorted list are then rendered pixel-by-pixel.
All pixels within a tile would first iterate over the Gaussians in the sorted list and perform $\alpha$-checking. 
$\alpha$-checking is used to filter out the Gaussians that do not intersect with the individual pixels.
A Gaussian is considered to intersect with a pixel if its computed transparency, $\alpha_i$, at that pixel exceeds a predefined threshold, $\alpha^*$.
Conceptually, each pixel forms its own list of intersected Gaussians after $\alpha$-checking, as shown in \Fig{fig:algo_overview}.

Once this is complete, each pixel would integrate the color contributions from its own list and accumulate the final pixel value.
This color integration process is governed by,
\begin{align}
\label{eqn:nerf}
   C(\textbf{p}) & = \sum_{i=1}^{N} \Gamma_i \alpha_i \textbf{c}_{i},\ \text{where} \ \Gamma_i = \prod^{i-1}_{j=1} (1-\alpha_j),
\end{align}
where $C(\mathbf{p})$ is the final color of pixel $\mathbf{p}$, $\alpha_i$ and $\mathbf{c}_i$ denote the transparency and color of the $i$th Gaussian, and $\Gamma_i$ represents the accumulated transmittance along the ray from the first Gaussian to the $(i-1)$th Gaussian. 
% The transparency $\alpha_i$ of each Gaussian is determined by its opacity $\sigma_i$ and its 2D covariance matrix~\cite{zwicker2001ewa}.
% $\textbf{c}_i$ is a function of the Gaussian spherical harmonics.

\paragraph{Backward Pass.}
Once the image is rendered, the backward pass first calculates the pixel-wise loss between the rendered image and the reference image.
This loss is then backpropagated to the relevant Gaussians to update their parameters.
Overall, the backward pass consists of two main stages: \textit{reverse rasterization} and \textit{re-projection}.

\subparagraph{Reverse Rasterization.}
Overall, this stage computes the partial gradients of every pixel-Gaussian pair and then aggregates the relevant gradients to individual Gaussians.

This process begins by identifying the intersected Gaussian list for each pixel via $\alpha$-checking.
Similar to the forward pass, $\alpha$-checking uses the previously cached tile-Gaussian sorted list from the forward pass to obtain the per-pixel intersection lists.

Given these per-pixel lists, the reverse color integration reverses the color integration process defined in \Eqn{eqn:nerf} and computes the partial gradients of individual Gaussians for each pixel.
Unlike forward color integration, the reverse color integration processes Gaussians from the $N$th Gaussian to the $1$st Gaussian.
The resulting gradients for each pixel-Gaussian pair are then collected into the pixel-Gaussian gradient list.

Using the pixel-Gaussian gradients, the aggregation stage collects all gradients associated with each Gaussian and calculates the accumulated gradient for each Gaussian.

\subparagraph{Re-Projection.}
This stage then transforms the accumulated gradients from the camera coordinate system to the world coordinate system.
This stage's computation is often lightweight.
% It is often merged into the aggregation stage.

\section{Motivation}
\label{sec:motiv}

We begin by showing the overall performance of 3DGS-SLAM algorithms.
We then examine the key factors that contribute to the main computational overheads.

\subsection{Performance Characteristics}
\label{sec:motiv:perf}

\begin{figure}[t]
\centering
\begin{minipage}[t]{0.48\columnwidth}
  \centering
  \includegraphics[width=\columnwidth]{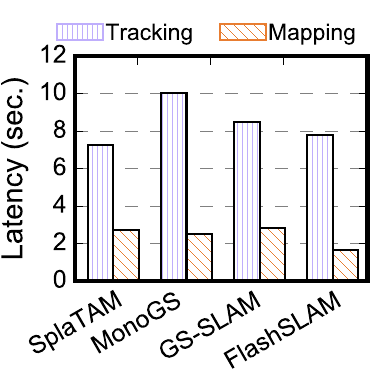}
  \caption{The amortized latency of \textit{tracking vs. mapping} across algorithms~\cite{keetha2024splatam, yugay2023gaussian, matsuki2024gaussian, pham2024flashslam}. Tracking dominates the execution.}
  \label{fig:latency}
\end{minipage}
\hspace{2pt}
\begin{minipage}[t]{0.48\columnwidth}
  \centering
  \includegraphics[width=\columnwidth]{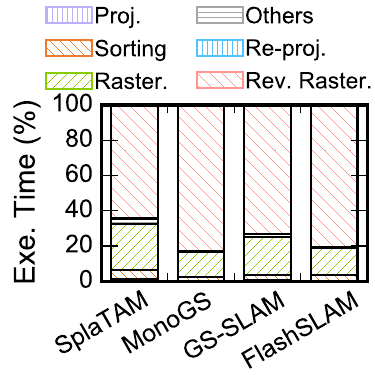}
  \caption{Normalized execution breakdown across algorithms. Rasterization and reverse rasterization dominate the execution.}
  \label{fig:exec_time}
\end{minipage}
\end{figure}

\paragraph{Latency.}
\Fig{fig:latency} shows the amortized per-frame latency of tracking and mapping across different 3DGS-SLAM algorithms on a Nvidia Ampere mobile GPU~\cite{orinsoc}. 
The results are averaged over all scenes in the Replica dataset~\cite{straub2019replica}. 
Tracking has much higher per-frame latency than mapping, because tracking is executed more frequently than mapping.
Thus, the latency of mapping can often be hidden behind tracking. 
% Overall, the amortized per-frame latency of tracking dominates the entire execution across algorithms.

\paragraph{Execution Breakdown.}
We further break down the execution time of tracking and mapping. 
Since both passes share the same pipeline in \Fig{fig:algo_overview}, \Fig{fig:exec_time} shows the execution time breakdown of key stages in the forward and backward passes. 
Across different algorithms, the primary execution bottleneck in the forward pass is rasterization, while reverse rasterization dominates the time of the backward pass. 
Together, these two stages account for 94.7\% of the execution time.
% This result is aligned with prior studies~\cite{wu2024gauspu, durvasula2025arc, he2025gsarch, feng2025lumina, lee2024gscore}.

\subsection{Performance Bottlenecks}
\label{sec:motiv:ch}

\begin{figure}[t]
\centering
\begin{minipage}[t]{0.48\columnwidth}
  \centering
  \includegraphics[width=\columnwidth]{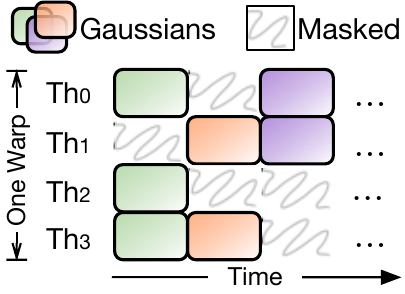}
  \caption{An example of warp divergence when different threads integrate Gaussians in color integration.}
  \label{fig:divergence_example}
\end{minipage}
\hspace{2pt}
\begin{minipage}[t]{0.48\columnwidth}
  \centering
  \includegraphics[width=\columnwidth]{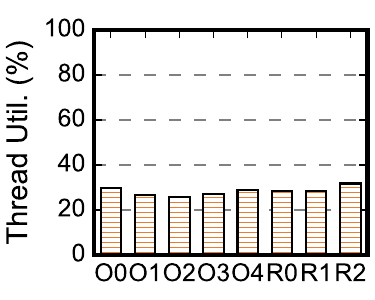}
  \caption{The thread utilization in rasterization is low. The x-axis shows different scenes from Replica~\cite{straub2019replica}.}
  \label{fig:thread_util}
\end{minipage}
\end{figure}

We further dissect the bottlenecks in forward and backward passes in the following characterizations, using SplaTAM~\cite{keetha2024splatam}.

\paragraph{Warp Divergence.}
As shown in \Fig{fig:algo_overview}, both projection and sorting in the forward pass are performed at the tile granularity to amortize computational cost across pixels within the same tile. 
However, rasterization must be executed at the pixel level to guarantee the rendering correctness: it first performs $\alpha$-checking to identify the contributing Gaussians for each pixel and then integrates only those Gaussians. 

In the current rasterization pipeline, each thread is responsible for rendering one pixel.
To leverage GPU parallelism, the color integration process broadcasts Gaussians within a GPU warp and masks those threads that do not need to integrate these Gaussians, as illustrated in \Fig{fig:divergence_example}.
Such a process would inevitably cause warp divergence.
We further measure GPU thread utilization during color integration, as shown in \Fig{fig:thread_util}. 
On average, thread utilization is only 28.3\%.
This shows that the warp divergence during rasterization is severe.

\begin{figure}[t]
\centering
\begin{minipage}[t]{0.48\columnwidth}
  \centering
  \includegraphics[width=\columnwidth]{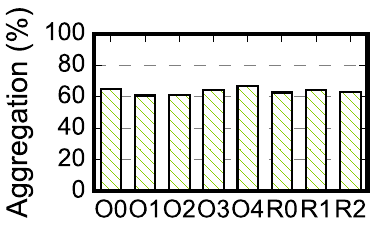}
  \caption{The execution percentage of aggregation in the reverse rasterization stage using Replica~\cite{straub2019replica} dataset.}
  \label{fig:atomic_add}
\end{minipage}
\hspace{2pt}
\begin{minipage}[t]{0.48\columnwidth}
  \centering
  \includegraphics[width=\columnwidth]{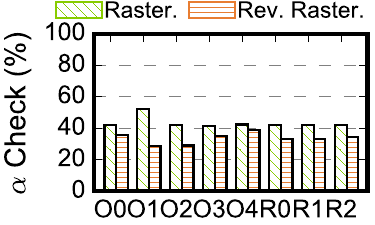}
  \caption{The execution percentage of $\alpha$-checking in both rasterization and reverse rasterization.}
  \label{fig:alpha_check}
\end{minipage}
\end{figure}

\paragraph{Aggregation.}
As prior studies~\cite{durvasula2025arc, wu2024gauspu, he2025gsarch} show, one main reason that reverse rasterization is a key bottleneck is that Gaussians must gather partial gradients from different pixels. 
To avoid race conditions, current GPU pipelines rely on \texttt{atomicAdd} operations during the aggregation stage of reverse rasterization. 
However, these atomic operations would lead to frequent pipeline stalls.
\Fig{fig:atomic_add} shows the aggregation overhead in reverse rasterization.
Over 63.5\% of its execution time is spent on aggregation. 
Thus, reducing data contention in aggregation is critical for improving its performance.

\paragraph{$\alpha$-Checking.}
Lastly, another reason that both rasterization and reverse rasterization are time-consuming is that $\alpha$-checking must be performed for every pixel-Gaussian pair.
Each $\alpha$-checking requires evaluating an expensive operation, the exponential function, which must be executed by special functional units (SFUs), rather than massive compute cores~\cite{ampere_arch}.
As shown in \Fig{fig:alpha_check}, $\alpha$-checking accounts for roughly 43.4\% and 33.6\% of the total time in rasterization and reverse rasterization, respectively.
Thus, reducing the number of $\alpha$-checks is critical to speed up these two stages.

% The root cause is the imbalanced compute units within modern GPUs. 
% For instance, in Nvidia Ampere GPUs, most operations can be parallelized by massive compute cores. 
% However, the computation of $\alpha$-checking requires exponential computation, which must be executed by special functional units (SFUs). 
% Since compute cores vastly outnumber the SFUs (16:1)~\cite{ampere_arch}, the exponential computation becomes a key bottleneck in the pipeline.

\section{Algorithm}
\label{sec:algo}

This section introduces our sparse processing framework, which reduces the overall computation by two orders of magnitude while maintaining task accuracy.
\Sect{sec:algo:sample} presents our adaptive pixel sampling tailored for tracking and mapping.
\Sect{sec:algo:pipeline} then introduces our \textit{pixel-based rendering} pipeline that fully unleashes the performance potential of sparse pixel processing.
Lastly, \Sect{sec:algo:analysis} revisits the remaining bottlenecks, which motivate our hardware support in \Sect{sec:arch}.

\subsection{Adaptive Pixel Sampling}
\label{sec:algo:sample}

In \Fig{fig:algo_overview}, the overall computation of both the forward and backward passes in 3DGS-SLAM is roughly proportional to the number of rendered pixels. 
To reduce the computation in 3DGS-SLAM, our sparse sampling algorithm aims to select a subset of pixels for processing.
Similar to prior studies~\cite{kodukula2021rhythmic, feng2022real, wang2025process, kravets2022progressive, egiazarian2007compressed}, our sampling algorithm also exploits task-specific characteristics of tracking and mapping to reduce pixel sampling rates without impacting accuracy.
However, prior work targets different domains, e.g., object detection\mbox{~\cite{kodukula2021rhythmic}} and eye tracking\mbox{~\cite{feng2024blisscam, feng2022real, wang2025process}}.
Their sampling techniques cannot be directly applied to 3DGS-SLAM.
Thus, we design a new sampling algorithm for SLAM in the following paragraphs.

\paragraph{Tracking.}
The optimization target of tracking is to estimate the pose of the current frame via an iterative training process. 
Inspired by traditional SLAM algorithms~\cite{zhang2014loam, newcombe2011kinectfusion, hess2016real}, which detect and match key feature points for localization, we observe that, for each frame, tracking in 3DGS-SLAM only requires optimizing a single camera pose, i.e., a $4 \times 4$ transformation matrix. 
Such a process should not necessitate using all pixels from a dense frame. 
Therefore, our algorithm selectively processes only the necessary pixels in tracking.

Unlike prior study~\cite{wu2024gauspu}, which performs sampling at the granularity of image tiles, our sampling algorithm operates at the \textit{pixel} level.
Specifically, our algorithm sparsely selects one pixel per $w_t \times w_t$ image tile.
Such a design has two main purposes: 1) adjacent pixels often carry similar information, tile-based selection would introduce redundant computation; 2) sampling one pixel per tile could capture the global features, making tracking more robust.

\begin{figure}[t]
\centering
\begin{minipage}[t]{0.48\columnwidth}
  \centering
  \includegraphics[width=\columnwidth]{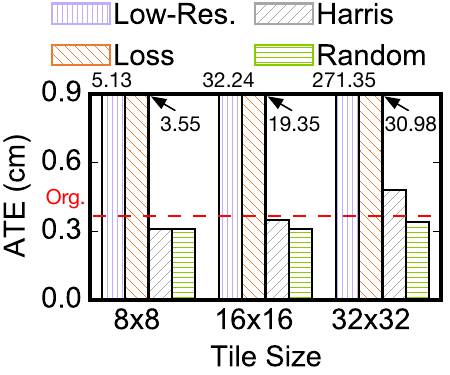}
  \caption{
  The SLAM tracking error with different sampling strategies and tile sizes during tracking, using SplaTAM~\cite{keetha2024splatam}.
  Here, lower is better. The red line shows the baseline accuracy.
  Random sampling achieves the best performance with robustness.
  }
  \label{fig:tracking_sampling}
\end{minipage}
\hspace{2pt}
\begin{minipage}[t]{0.48\columnwidth}
  \centering
  \includegraphics[width=\columnwidth]{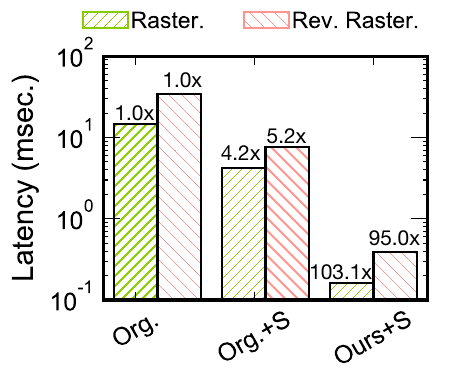}
  \caption{The execution latency of rasterization and reverse rasterization on a mobile Ampere GPU~\cite{orinsoc} during tracking, using SplaTAM~\cite{keetha2024splatam}.
  ``S'': applying sparse pixel sampling.
  Numbers represent speedups compared to the original pipeline.
  }
  \label{fig:bottleneck_breakdown}
\end{minipage}
\end{figure}

\Fig{fig:tracking_sampling} shows the tracking accuracies of different sampling strategies under the same sampling rate. 
Here, ``Low-Res.'' stands for downsampling to low-resolution images, while ``Loss'' stands for the method from GauSPU~\cite{wu2024gauspu}.
Both ``Random'' and ``Harris'' apply our sampling strategy with different selection metrics.
``Random'' select one pixel from $ w_t \times w_t $ tile randomly, while ``Harris'' selects based on Harris descriptor~\cite{harris1988combined}.
Overall, methods lacking global coverage, e.g., GauSPU~\cite{wu2024gauspu} or simply reducing the resolution, lead to low accuracy.
In comparison, random sampling achieves equivalent or better accuracy compared to feature-based methods.
For simplicity, we choose random sampling for tracking.

% In \Fig{fig:tracking_sampling}, at higher sampling rates (e.g., one pixel per $8 \times 8$ tile), feature-based methods, such as Harris descriptor~\cite{harris1988combined}, yield the best accuracy. 
% However, under extremely low sampling rates (e.g., one pixel per $32 \times 32$ tile), uniform sampling achieves comparable accuracy with better computational efficiency.

\begin{figure}[t]
    \centering
    \includegraphics[width=\columnwidth]{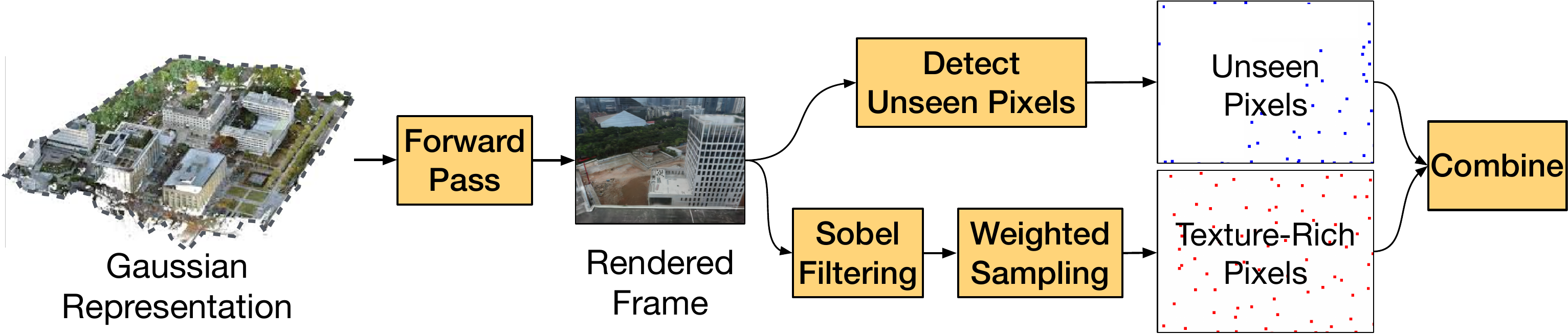}
    \caption{The sampling algorithm for mapping. Two types of pixels are sampled in mapping. 
    The first type is the pixels that were unseen in previous reconstructions, allowing the algorithm to focus on refining newly observed regions. 
    The second type includes texture-rich pixels sampled across the entire image to capture global structural information.}
    \label{fig:mapping_sampling_algo}
\end{figure}

\paragraph{Mapping.}
Unlike tracking, which is used to estimate poses, the purpose of mapping is to reconstruct the previously unseen regions of the scene.
In \Fig{fig:mapping_sampling_algo}, we design our sampling algorithm to prioritize the unseen pixels.
% , i.e., unexplored regions or previously occluded areas.
These pixels are typically concentrated along previously occluded object boundaries, where depth variations are significant, or in regions that were previously unexplored.
At a given timestamp $t$, we define whether a pixel, $\mathbf{p}$, in the current image frame is unseen based on its accumulated transmittance, $\Gamma_{\text{final}}(\mathbf{p})$,
\begin{align}
    \cF(\mathbf{p}) = \begin{cases}
       \text{unseen},\ \text{if}\ \Gamma_{\text{final}}(\mathbf{p})\ >\ 0.5, \\
       \text{seen},\ \text{otherwise}.                           \\
    \end{cases}                                 
\end{align}
Here, $\Gamma_{\text{final}}(\mathbf{p})$ is computed during the first forward pass (we perform only once per mapping). 
A high transmittance means that very few Gaussians have contributed to this pixel.
Therefore, function $\cF(\mathbf{p})$ selects those pixels to better improve regions that still require reconstruction.

However, selecting only unseen pixels often leads to poor tracking accuracy, as such pixels tend to be too sparse (see results in \Sect{sec:eval:abl}). 
Thus, in addition to the unseen pixels, we also sample additional pixels across the image using a weighted sampling strategy, in \Fig{fig:mapping_sampling_algo}.
Specifically, we assign texture-rich pixels with higher probability and select one pixel per $w_m \times w_m$ tile.
The probability, $\cP(\mathbf{p})$, is defined as follows, 
\begin{equation}
    \cP(\mathbf{p}) = w_R(\mathbf{p}) \times r,\ \text{where}\ w_{R}(\mathbf{p}) = \sqrt{G_{\text{x}}^2+G_{\text{y}}^2},
\end{equation}
where $G_{\text{x}}$ and $G_{\text{y}}$ are the horizontal and vertical gradients at pixel $\mathbf{p}$ using Sobel filters~\cite{kanopoulos1988design}. 
$w_{R}$ stands for the overall gradient, which approximates the local texture richness.
$r$ is a random floating number between 0 and 1.
\Sect{sec:eval:abl} further compares our strategy against other methods.

\begin{figure*}
    \centering
    \includegraphics[width=\textwidth]{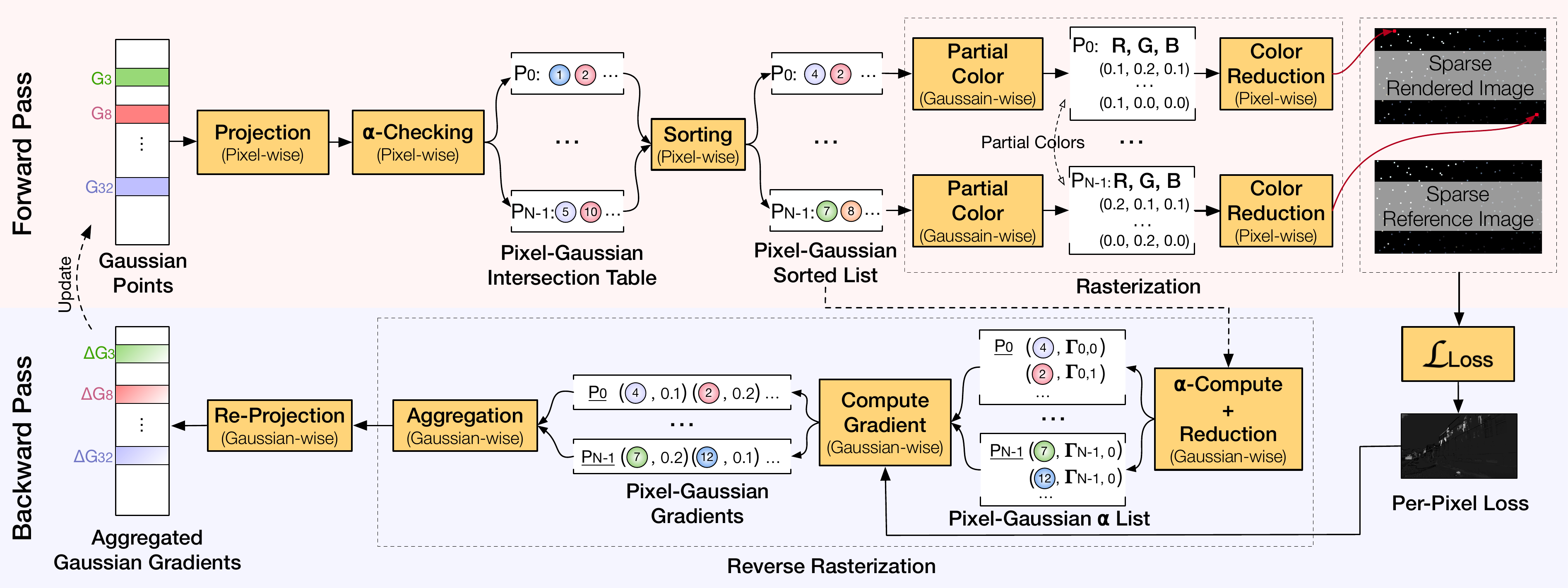}
    \caption{
    Overview of our \textit{pixel-based rendering} pipeline for sparse 3DGS-SLAM.
    To improve GPU thread utilization, we shift from pixel-wise parallelism to Gaussian-wise parallelism in both the rasterization and reverse rasterization stages. 
    Instead of assigning one thread per pixel, our pipeline enables threads within a GPU warp to co-render a single pixel.
    Additionally, we introduce an optimization, preemptive $\alpha$-checking, that moves $\alpha$-checking from rasterization to projection in the pipeline. 
    This not only reduces the workload of subsequent stages but also eliminates warp divergence in rasterization.
    }
    \label{fig:our_pipeline}
\end{figure*}

\subsection{Pixel-Based Rendering}
\label{sec:algo:pipeline}

\paragraph{Motivation.}
While our sparse sampling algorithm largely reduces the number of processed pixels during both the forward and backward passes, we find that its actual speedup on existing mobile GPUs is limited. 
In \Fig{fig:bottleneck_breakdown}, we test a case where we process one pixel per $16 \times 16$ tile on a mobile Ampere GPU~\cite{orinsoc}, we expect a 256$\times$ speedup on the two dominated stages: rasterization and reverse rasterization, since their computations are proportional to the number of pixels.

However, \Fig{fig:bottleneck_breakdown} shows that applying our sampling algorithm to the original SplaTAM (denoted as ``Org.+S'') yields only 4.2$\times$ and 5.2$\times$ speedup in rasterization and reverse rasterization, respectively, far below the theoretical gain.
The root cause of this gap is the thread-to-pixel mapping in the original 3DGS pipeline, where each GPU thread is responsible for rendering a single pixel. 
Under sparse sampling, only one thread in a GPU warp does meaningful work, while other threads are idle, resulting in severe PE under-utilization.

\paragraph{Pipeline.}
To address this issue, we propose a new rendering paradigm, \textit{pixel-based rendering} in \Fig{fig:our_pipeline}, that parallelizes rasterization and reverse rasterization at the \textit{Gaussian} level.
By doing so, our pipeline significantly mitigates the PE under-utilization introduced by the original pipeline.

\underline{In the forward pass}, we make the two key changes.

\circled{white}{1} 
We process projection and sorting at the pixel level, rather than at the tile level as in the original pipeline.
Since our algorithm requires only a sparse subset of pixels, tile-level projection, i.e., identifying Gaussians that intersect with an entire tile, introduces redundant computations by including Gaussians that do not intersect with the sampled pixels.
In contrast, performing the projection stage at the pixel level ensures that only the intersected Gaussians would be included in the subsequent stages, eliminating unnecessary work.

\circled{white}{2} 
We redesign the rasterization stage to operate at the granularity of Gaussians, such that each pixel is co-rendered by a warp of threads. 
Specifically, all threads within a warp share the same Gaussian list associated with a target pixel. 
The Gaussian list is then evenly distributed across the threads in the warp to ensure an even workload across threads.
During rasterization, each thread computes the partial colors from a subset of Gaussians, as illustrated in \Fig{fig:our_pipeline}.
Once partial colors are computed, a reduction operation is performed to accumulate the final color of that pixel.
% This way not only improves warp utilization but also achieves better workload balance across threads.

\underline{In the backward pass}, the primary change is on the reverse rasterization stage.
Similar to the forward rasterization, we shift from pixel-level parallelism to Gaussian-level parallelism.
Unlike forward rasterization, our reverse rasterization inverts the color integration process, with two rounds of reductions.
The first round is introduced by our pixel-based rendering.

\circled{white}{1} 
In the first reduction, we need first to compute the individual Gaussian transparency, $\alpha_i$, in parallel across threads, and then perform a cross-thread reduction to obtain $\Gamma_i$ for each Gaussian by accumulating the transparencies $\prod^{i-1}_{j=1} (1-\alpha_j)$ of all preceding Gaussians as defined in \Eqn{eqn:nerf}. 
The original pipeline does not require this reduction because $\Gamma_i$ values are computed and accumulated sequentially by a single thread.

\circled{white}{2} 
Once $\Gamma_i$ is computed, each thread can then independently calculate the partial gradients from the pixel to its relevant Gaussians.
This step is also fully parallelizable in \Fig{fig:our_pipeline}.
Lastly, a second reduction is required to aggregate all partial gradients associated with each Gaussian, similar to the original pipeline.
Although the second reduction involves atomic operations that cause data contentions, luckily, our sparse processing naturally alleviates these data contentions.

\paragraph{Preemptive $\alpha$-Checking.}
In addition to pixel-based rendering, we also introduce \textit{preemptive $\alpha$-checking}, an optimization that eliminates the warp divergence in rasterization. 
In the original pipeline, $\alpha$-checking is needed during rasterization to filter out Gaussians that do not intersect with the target pixel or whose transparency is below a threshold, $\alpha^*$. 
Since different pixels often integrate different subsets of Gaussians, $\alpha$-checking would introduce warp divergence (\Fig{fig:divergence_example}).
% as shown in \Fig{fig:divergence_example}.

In our pipeline, the Gaussian list in rasterization is no longer shared across pixels, which enables us to move $\alpha$-checking earlier in the pipeline.
This way, we can effectively pre-filter Gaussians that will have no contribution to the target pixel. 
As a result, we reduce the workloads of subsequent sorting and rasterization.
More importantly, the pixel-Gaussian sorted list in \Fig{fig:our_pipeline} contains only relevant Gaussians that need to be integrated into the final pixel value, ensuring no warp divergence in rasterization and reverse rasterization.

\paragraph{Walk-Through.}
\Fig{fig:our_pipeline} illustrates the execution flow of our algorithm.
In the forward pass, we first perform projection and $\alpha$-checking for individual pixels to construct an intersection table that records the set of intersected Gaussians for each pixel. 
Next, the intersected Gaussians are sorted by depth for each pixel. 
Once the pixel-Gaussian sorted list is obtained, threads within a warp collaboratively compute the partial colors of all Gaussians in parallel. 
Finally, we perform a color reduction to integrate partial colors into the final pixel.

% In the backward pass, the rendered sparse image is first compared against the reference image to compute the per-pixel loss. 
In the backward pass, we first compute the per-pixel loss. 
We then reuse the pixel-Gaussian sorted list from the forward pass to compute the transparency $\alpha_i$ of each intersected Gaussian in parallel. 
Next, each intersected pixel-Gaussian pair computes the accumulated transmittance $\Gamma_i$ by accumulating the $\alpha_i$ values of all preceding Gaussians via across-thread reductions (\Eqn{eqn:nerf}). 
Using the computed $\Gamma_i$ values, each thread then calculates the partial gradient of each intersected pixel-Gaussian pair.
A second reduction is then used to aggregate the partial gradients for each Gaussian. 
Finally, the aggregated gradients are backpropagated to Gaussians via re-projection.

% \begin{figure}[t]
% \centering
% \begin{minipage}[t]{0.48\columnwidth}
%   \centering
%   \includegraphics[width=\columnwidth]{projection_bottleneck}
%   \caption{The execution latency (left y-axis) and the relative forward time (right y-axis) of projection on mobile Ampere GPU during tracking. Projection becomes the new bottleneck in the forward pass.}
%   \label{fig:projection_bottleneck}
% \end{minipage}
% \hspace{2pt}
% \begin{minipage}[t]{0.48\columnwidth}
%   \centering
%   \includegraphics[width=\columnwidth]{reverse_rasterization_bottleneck}
%   \caption{The execution latency (left y-axis) and the relative backward time (right y-axis) of reverse rasterization on mobile GPU during tracking. Reverse rasterization remains the key bottleneck in backward.}
%   \label{fig:reverse_rasterization_bottleneck}
% \end{minipage}
% \end{figure}

\begin{figure}[t]
    \centering
    \includegraphics[width=\columnwidth]{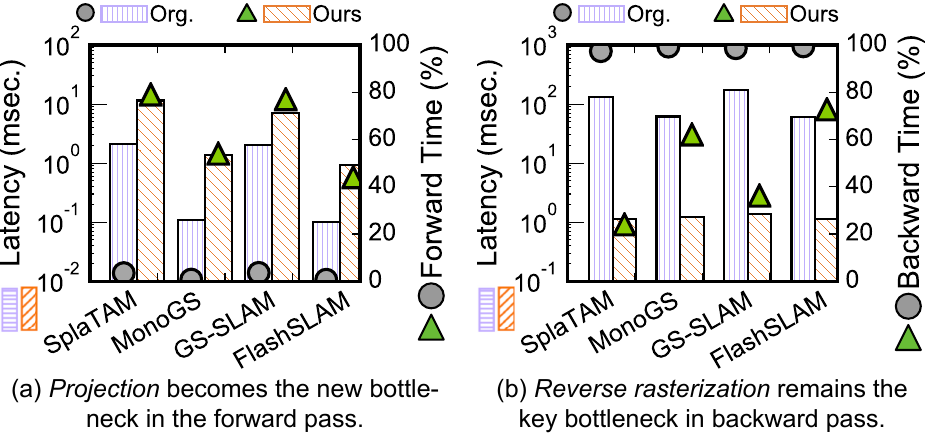}
    \caption{The execution latency (left y-axis) and the relative time (right y-axis) of \textit{projection} and \textit{reverse rasterization} on mobile GPU during tracking.}
    \label{fig:proj_raster}
\end{figure}

\subsection{Bottleneck Analysis}
\label{sec:algo:analysis}

With all the optimizations together in \Sect{sec:algo:pipeline}, \Fig{fig:bottleneck_breakdown} shows that our pixel-based rendering achieves 103.1$\times$ and 95.0$\times$ speedup on rasterization and reverse rasterization on SplaTAM~\cite{keetha2024splatam}.
However, by further analyzing the execution time breakdown of our pixel-based rendering pipeline, our experiment shows that the performance bottleneck shifts from rasterization to projection in the forward pass, as shown in \Fig{fig:proj_raster}a. 
The proportion of forward time spent in projection increases from 2.1\% to 63.8\%.
The main reason is that we move $\alpha$-checking to the projection stage, which substantially increases the workload of the projection stage.

Meanwhile, in the backward pass, although the overall execution time is reduced significantly, \Fig{fig:proj_raster}b shows that the relative backward time spent in reverse rasterization decreases from 98.7\% to 48.76\%.
However, reverse rasterization still accounts for the majority of the backward time (up to 72.7\%).
There are primarily two reasons: 1) our reverse rasterization introduces an additional round of across-thread reductions, which incurs synchronization overhead; 2) aggregation remains a key bottleneck due to frequent pipeline stalls caused by atomic operations, despite that sparse pixel processing helps to alleviate this data contention across threads.

\begin{figure*}
    \centering
    \includegraphics[width=\textwidth]{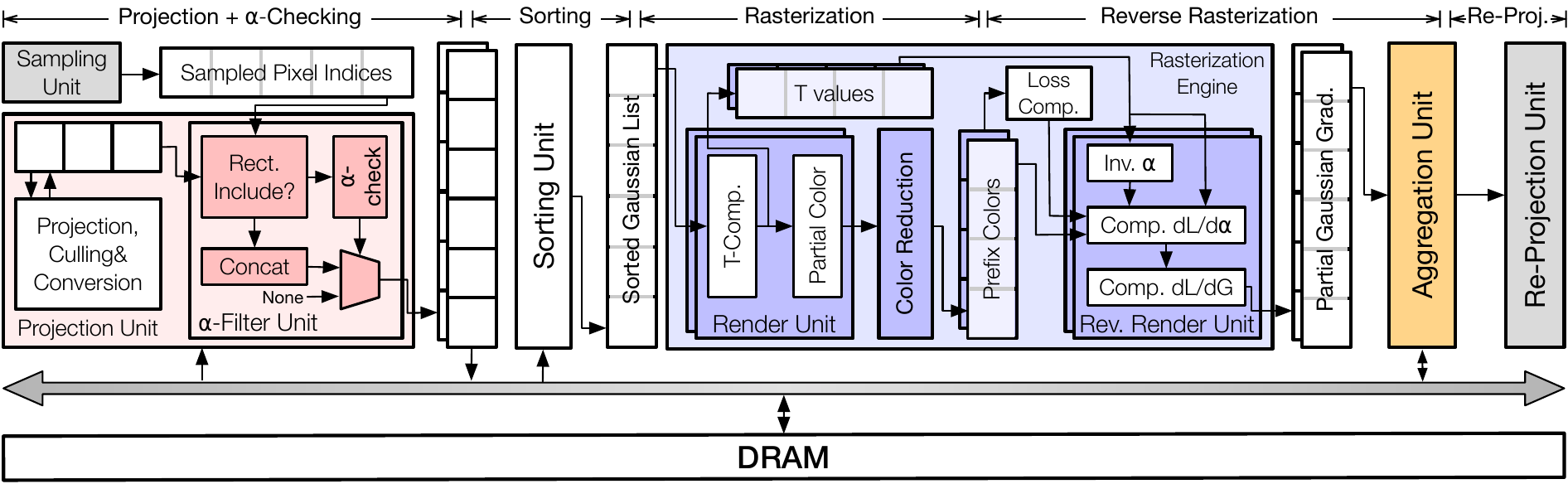}
    \caption{
    Overview of our pipelined architecture. Our architecture is built upon \textsc{MetaSapiens}~\cite{lin2025metasapiens}, a 3DGS accelerator for the forward pass only. 
    We augment the baseline architecture to support the backward pass. 
    Specifically, we co-design a simplified rasterization engine (purple-colored) that mitigates the PE under-utilization in rasterization and reverse rasterization. 
    We also propose a caching technique between these two stages to avoid the across-thread reduction in reverse rasterization (\Sect{sec:arch:raster}). 
    Meanwhile, we augment the projection unit (pink-colored) to support preemptive $\alpha$-checking and design a dedicated aggregation unit (yellow-colored) to accelerate reverse rasterization (\Sect{sec:arch:issues}).
    }
    \label{fig:arch}
\end{figure*}

\section{Architectural Support}
\label{sec:arch}

To address the bottleneck shifts brought by our pixel-based rendering, we propose our accelerator, \proj, to further boost the performance. 
We first give an overview (\Sect{sec:arch:overview}) and discuss how we design a lightweight pipelined architecture (\Sect{sec:arch:raster}) and address the key bottlenecks (\Sect{sec:arch:issues}).

\subsection{Overview}
\label{sec:arch:overview}

Following the philosophy of our pixel-based rendering, we design a dedicated architecture, as shown in \Fig{fig:arch}.
Our design builds upon \textsc{MetaSapiens}~\cite{lin2025metasapiens}, a pipelined 3DGS accelerator for the forward pass only.
We extend \textsc{MetaSapiens} to support both forward and backward passes, with the colored components highlighting our augmentations.

Specifically, we make the following contributions.
First, we design a streaming architecture for sorting, rasterization, and reverse rasterization stages, and enable the pipelining across those stages.
Second, to support pipelining, we propose a lightweight rasterization engine in {\Sect{sec:arch:raster}} that removes the redundant {$\alpha$}-checking components that are targeted for tile-based rendering and addresses the synchronization overhead in reverse rasterization, i.e., the first round of across-thread reduction due to pixel-based rendering.
Lastly, to address the bottleneck shifts described in {\Sect{sec:algo:analysis}}, we further augment our architecture with: 1) a projection unit to address the increased workload in projection and accelerate the preemptive {$\alpha$}-checking, and 2) an aggregation unit to alleviate the pipeline stalls due to data contention and frequent off-chip traffic.

\subsection{Rasterization Engine}
\label{sec:arch:raster}

In this section, we present our rasterization engine, which simplifies the core computation in prior designs~\cite{he2025gsarch, wu2024gauspu} and enables high parallelism across Gaussians.
Overall, the rasterization engine contains two sets of processing units: render units for rasterization and reverse render units for reverse rasterization.
We next describe them individually. 

\paragraph{Render Unit.}
One key inefficiency in prior render unit designs~\cite{lee2024gscore, lin2025metasapiens} is that each Gaussian undergoes $\alpha$-checking to determine if it contributes to a pixel.
This often leads to PE under-utilization on accelerators~\cite{feng2025lumina}.
To address this, our architectural design adopts the preemptive $\alpha$-checking from \Sect{sec:algo:pipeline}, which moves the $\alpha$-checking logic from rasterization to projection.
By doing so, we guarantee that only Gaussians that contribute to the target pixel proceed to rasterization.
This optimization allows us to skip the $\alpha$-checking logic from render units and compute the partial color of each Gaussian directly, as shown in \Fig{fig:arch}.
During rasterization, multiple render units read different Gaussian data and are executed in parallel.
A dedicated reduction unit then gathers these partial colors to accumulate the final pixel value.

\begin{figure}[t]
    \centering
    \includegraphics[width=\columnwidth]{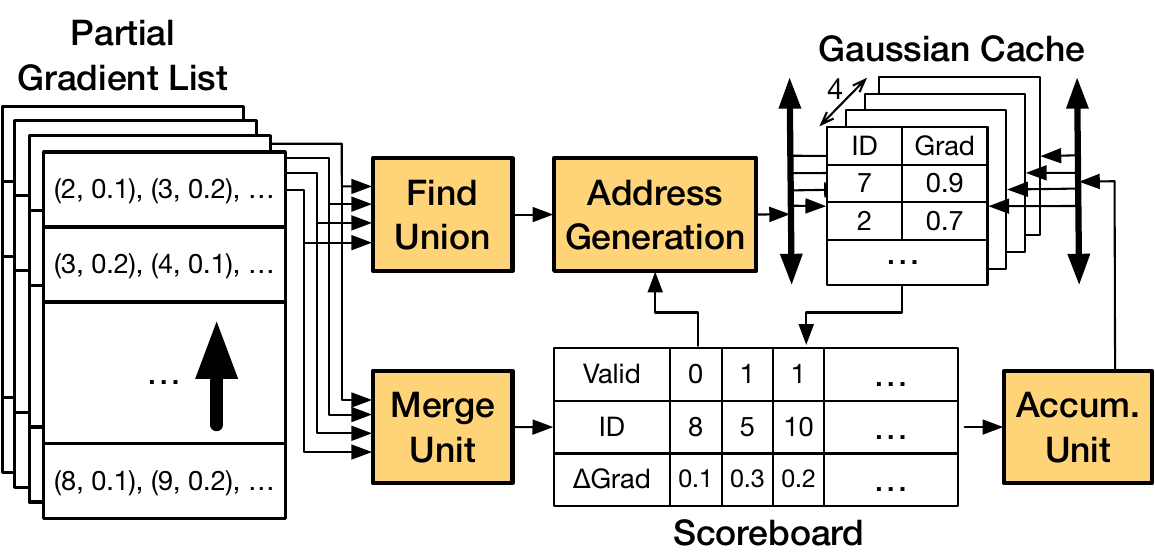}
    \caption{The design of the aggregation unit. We batch process the partial gradients of multiple pixels. A merge unit is to perform on-chip gradient reduction before accumulating with partial accumulated gradients from Gaussian cache. A scoreboard records which Gaussian is ready for accumulation.}
    \label{fig:aggregation_unit}
\end{figure}

\paragraph{Reverse Render Unit.}
\Sect{sec:algo:analysis} shows that, after shifting to pixel-based rendering, one overhead in reverse rasterization is the need to perform one additional round of across-thread reductions.
The additional round of reduction is to compute the accumulated transmittance $\Gamma_i$ for each Gaussian $i$ and the prefix color $C_i$, which integrates the partial colors from the first to the $i$th Gaussian.
Note that, both ${\Gamma_i}$ and ${C_i}$ are intermediate results during forward rasterization.
However, caching these intermediate values in the original tile-based rendering is prohibitively expensive.
Because every pixel in a dense frame requires storing these two values for all contributed Gaussians.
However, in our pixel-based rendering pipeline, we only need to store the data for one single pixel at a time, which is largely manageable and can be held entirely on-chip.

By exploiting this feature, we co-design the reverse render unit with the render unit and color reduction unit in the forward pass to output and store the intermediate values, i.e., $\{\Gamma_i\}$ and $\{C_i\}$, for one pixel in the on-chip buffer, as shown in \Fig{fig:arch}.
These cached data are then forwarded directly to the reverse render units.
Using these data, the reverse render units avoid implementing additional logic for reduction and inter-PE communication.
More importantly, eliminating these reduction operations removes inter-PE dependencies, so that the partial gradients for each Gaussian can be computed in parallel.
Once computed, the partial gradients are buffered on-chip for the aggregation stage, which is introduced in \Sect{sec:arch:issues}.

\subsection{Addressing Key Bottlenecks}
\label{sec:arch:issues}

\paragraph{Aggregation Unit.}
To support accumulating partial gradients during reverse rasterization, we introduce a dedicated aggregation unit following the rasterization engine.
In the backward pass, all partial gradients in the pixel-Gaussian gradient list need to be accumulated into their corresponding Gaussians.
However, the total number of partial gradients 
(roughly the product of the number of Gaussians and the number of pixels) 
is too large to perform on-chip reduction.
Moreover, the highly irregular accumulation patterns require frequently reloading unfinished accumulated gradients from off-chip memory. 
Thus, we design an aggregation unit to hide frequent pipeline stalls introduced by off-chip memory traffic.

\Fig{fig:aggregation_unit} shows the design of our aggregation unit.
Here, each entry in the partial gradient list contains the partial gradients from a single pixel.
During aggregation, the aggregation unit first reads the Gaussian IDs from $n$ entries, i.e., $n=4$ in \Fig{fig:aggregation_unit}.
We compute the union of Gaussian IDs across these entries and then load the corresponding accumulated gradients from off-chip memory into the Gaussian cache.

In parallel, the aggregation unit reads $n$ Gaussian tuples, i.e., the Gaussian ID and its partial gradient, and forwards them to the merge unit.
The merge unit performs intra-batch reductions by combining gradients with identical Gaussian IDs and stores the results in the scoreboard.
The role of the scoreboard is to maintain the current status of stored Gaussians: their IDs and whether their partial accumulated gradients have been loaded into the cache.
The partial accumulated gradients mean those accumulated Gaussian gradients that are done partially and are unfinished.
Meanwhile, each cycle, the accumulation unit checks the scoreboard for ready Gaussians, i.e., those whose partial accumulated gradients are available in the cache.
Then, the accumulation unit reads the partial accumulated gradients from the Gaussian cache, updates them with the partial gradients stored in the scoreboard, and writes back to the cache.
This way, we can effectively hide the off-chip traffic latency by simultaneously updating other Gaussian gradients.

\paragraph{Projection Unit.}
In \Sect{sec:algo:analysis}, we show that once applying our pixel-based rendering, the bottleneck in the forward pass shifts from rasterization to projection due to preemptive $\alpha$-checking (\Fig{fig:proj_raster}a).
Such a shift comes from two factors: 
1) each Gaussian’s bounding box (BBox) must be checked against all sampled pixels; and
2) $\alpha$-checking involves expensive exponential computations, as mentioned in \Sect{sec:motiv:ch}.

To address the first issue, we propose a direct indexing method using the four corners of each Gaussian’s BBox to limit the number of pixels we iterate.
Since our sampling algorithm selects one pixel per tile, we can directly compute the index range in the sampled pixel list using the minimal and maximal coordinates.
% The pixel index, $i_\text{idx}$, can be expressed as,
% \begin{equation}
%     i_\text{idx} =\lfloor\frac{(x_\text{max}-x_\text{min})}{w} \rfloor * \lfloor\frac{y}{w}\rfloor,\ \text{where}\ y \in [y_\text{min}, y_\text{max})
% \end{equation}
% where $\langle x_\text{min}, y_\text{min}\rangle$ and $\langle x_\text{max}, y_\text{max}\rangle$ are the minimum and maximum coordinates of the BBox, $w$ is the tile size in our algorithm.
This way, we avoid exhausting the entire pixel list and avoid unnecessary $\alpha$-checking.
Note that, the unseen pixel indices in mapping are stored separately, so that the unseen pixel indices do not interrupt our indexing strategy.
Second, to mitigate the computational cost of exponentiation, we approximate the exponential function with a lookup table (LUT)~\cite{liu2025voyager}.
Our empirical evaluation shows that a LUT with a size of 64 entries is sufficient to maintain the same accuracy.
% The sensitivity of accuracy to LUT size is shown in \Sect{sec:eval:sens}.
\section{Experimental Setup}
\label{sec:exp}

% \paragraph{GPU Implementation.}

\paragraph{Hardware Configuration.}
Overall, \proj has a basic configuration similar to \textsc{MetaSapiens}~\cite{lin2025metasapiens}.
\proj consists of eight projection units, four hierarchical sorting units, four rasterization engines, and one aggregation unit.
We augment each projection unit with four $\alpha$-filter units.
Each rasterization engine has $2 \times 2$ render units and $2 \times 2$ reverse render units with one color reduction unit in between.
To store the intermediate $\Gamma_i$ and $C_i$ values, each rasterization engine is designed with an 8~KB double buffer.
In addition, a 64~KB global double buffer is used to hold the intermediate data of the entire pipeline.
Lastly, the aggregation unit is designed with four channels to process the partial gradients of four pixels in parallel with a 32~KB Gaussian cache and a 8~KB scoreboard.

\paragraph{Experimental Methodology.}
For GPU performance, we measure latency, including the execution time as well as the kernel launch on the mobile Ampere GPU. 
The GPU power is directly obtained using the built-in power sensing circuitry on Orin.
For accelerator performance, we develop a RTL implementation of our pipelined architecture.
The design is clocked at 500 MHz.
The RTL design is implemented via Synposys
synthesis and Cadence layout tools in TSMC 16nm FinFET
technology.
The numbers of our RTL design are then scaled down to 8 nm node using DeepScaleTool~\cite{stillmaker2017scaling, sarangi2021deepscaletool} to match the mobile Ampere GPU on Nvidia Orin SoC in 8 nm node~\cite{orinsoc}.
The SRAMs are generated using the Arm Artisan memory compiler. 
% Power is estimated using Synopsys PrimeTimePX with annotated switching activities.
The DRAM is modeled after 4 channels of Micron 16 Gb LPDDR3-1600 memory~\cite{micronlpddr3}. 
The DRAM energy is obtained using Micron System Power Calculators~\cite{microdrampower}.

\paragraph{Area.} Overall, \proj has a smaller area (1.07~mm$^2$) compared to other 3DGS accelerators, such as GSCore (1.77~mm$^2$)~\cite{lee2024gscore} and GSArch (3.42~mm$^2$)~\cite{he2025gsarch}, with all areas scaled down to 16nm node using DeepScaleTool~\cite{stillmaker2017scaling, sarangi2021deepscaletool}.
The primary contributor to \proj compact design is its efficient rasterization engine, which accounts for only 28\% of the total area.
The remaining stages occupy 57\% of the area.
Some are due to the larger projection units.
The rest are SRAMs, which comprise 15\% of the area.

\paragraph{Software Setup.}
We evaluate \mbox{\proj} on the two widely used indoor SLAM datasets, Replica{~\cite{straub2019replica}} and TUM RGB-D{~\cite{sturm2012benchmark}}.
Replica comprises eight sequences.
Each sequence consists of 2000 RGB-D images.
TUM RGB-D is a more complex real-world dataset with fast camera motion.
To evaluate the effectiveness of our sampling algorithms, we use four different 3DGS-SLAM algorithms: SplaTAM~\cite{keetha2024splatam}, MonoGS~\cite{matsuki2024gaussian}, GS-SLAM~\cite{yugay2023gaussian}, and FlashSLAM~\cite{pham2024flashslam}.

\paragraph{Metrics.} 
We report two standard accuracy metrics: absolute trajectory error (ATE), which is used to measure the accuracy of pose estimation, and peak signal-to-noise ratio
(PSNR) which is to measure the reconstruction quality.

\paragraph{Baselines.}
We compare three hardware baselines:
\begin{itemize}
    \item \mode{GPU}: a mobile Ampere GPU on Nvidia Orin SoC~\cite{orinsoc}.
    \item \mode{GSArch}~\cite{he2025gsarch}: a dedicated 3DGS training architecture for the tile-based rendering pipeline. 
    Here, we compare the edge configuration reported in the GSArch paper.
    \item \mode{GauSPU}~\cite{wu2024gauspu}: a dedicated accelerator for 3DGS-SLAM. It executes projection and sorting on GPU, and the remaining stages are executed on the dedicated accelerator.
    Here, we model the GPU performance using the parameters obtained from the mobile GPU on Orin.
\end{itemize}
To ensure a fair comparison with the mobile GPU performance on Orin, we scale both designs down to 8~nm node using DeepScaleTool~\cite{stillmaker2017scaling, sarangi2021deepscaletool} and clock them at 500~MHz.
We do not include inference-only 3DGS accelerators~\cite{feng2025lumina, ye2025gaussian, lee2024gscore, li2025uni, lee2025vr, lin2025metasapiens} in our evaluation, because their delay will be dominated by the backward pass latency (see \Fig{fig:exec_time}).

\paragraph{Variants.} 
We evaluate two variants of \proj to separate the contributions in our paper: \mode{\proj-sw}, which executes our pixel-based rendering on mobile GPUs; and \mode{\proj-hw}, which executes our pixel-based rendering on our proposed pipelined architecture.
\section{Evaluation}
\label{sec:eval}

\subsection{Accuracy}
\label{sec:eval:acc}

% \begin{figure}[t]
% \centering
% \begin{minipage}[t]{0.48\columnwidth}
%   \centering
%   \includegraphics[width=\columnwidth]{eval_ate}
%   \caption{The tracking accuracy comparison between the original algorithms and our sampling algorithm. Lower is better.}
%   \label{fig:eval_ate}
% \end{minipage}
% \hspace{1pt}
% \begin{minipage}[t]{0.48\columnwidth}
%   \centering
%   \includegraphics[width=\columnwidth]{eval_psnr}
%   \caption{The reconstruction quality comparison between the original algorithms and our sampling algorithm. Higher is better.}
%   \label{fig:eval_psnr}
% \end{minipage}
% \end{figure}

\begin{figure}[t]
    \centering
    \includegraphics[width=\columnwidth]{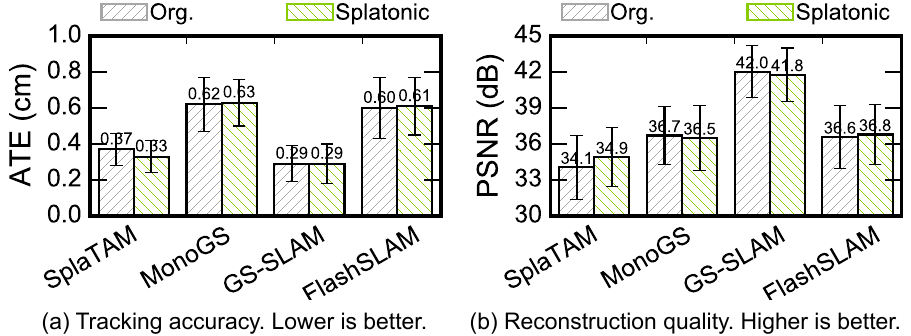}
    \caption{The tracking accuracy and reconstruction quality comparison between the baselines and our sampling algorithm across 8 sequences on Replica.}
    \label{fig:overall_acc}
\end{figure}

\begin{figure}[t]
    \centering
    \includegraphics[width=\columnwidth]{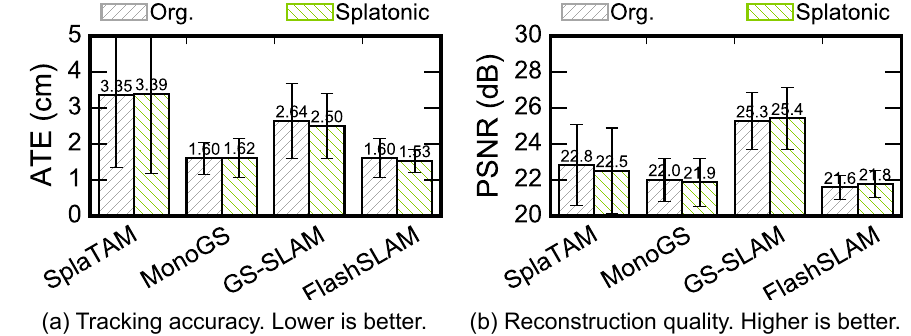}
    \caption{The tracking accuracy and reconstruction quality comparison between the baselines and our sampling algorithm over 3 sequences on TUM RGB-D.}
    \label{fig:overall_tum_acc}
\end{figure}

In our sampling algorithm, we set the tile size to $w_t = 16 \times 16$ for tracking and $w_m = 4 \times 4$ for mapping. During mapping, we perform one full-frame mapping for every four frames. 
Unless otherwise specified, this configuration is used as the default setting for the remaining evaluations.

\paragraph{Tracking Accuracy.} {\Fig{fig:overall_acc}}a and {\Fig{fig:overall_tum_acc}}a show the tracking accuracy comparison between the original 3DGS-SLAM algorithms and the ones with our sparse sampling algorithm the tracking accuracy.
Overall, \proj matches or outperforms the performance of original algorithms.
Across four different algorithms, the average ATEs of \mbox{\proj} are 0.46 cm and 2.26 cm on Replica and TUM RGB-D, respectively.
\mbox{\proj} are 0.01 and 0.03 lower than the baselines.
\Review{C2} 
Our better accuracy stems from the fact that sparsely sampled pixels can help eliminate the false matching in regions with repetitive patterns, similar to extracting keypoint descriptors in conventional SLAM. 
The same principle in conventional SLAM is also applied to 3DGS-SLAMs.

\paragraph{Reconstruction Quality.} {\Fig{fig:overall_acc}}b and {\Fig{fig:overall_tum_acc}}b compare the reconstruction quality between the original algorithms and the ones with our proposed sparse sampling approach. 
Across all four 3DGS-SLAM algorithms, \proj achieves higher PSNR values on average. 
Notably, \proj outperforms the baseline by 0.8~dB on SplaTAM~\cite{keetha2024splatam}. 
This demonstrates that our sampling strategy in the mapping process effectively directs the training focus toward unseen regions and texture-rich areas, thus enhancing the overall reconstruction quality.

\subsection{GPU Performance.}
\label{sec:eval:gpu}

\begin{figure}[t]
    \centering
    \includegraphics[width=\columnwidth]{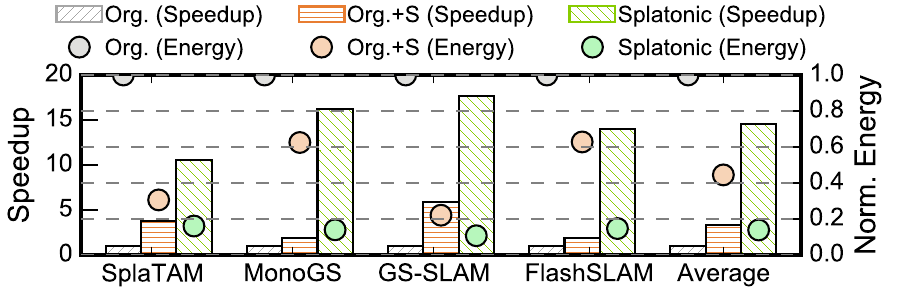}
    \caption{The end-to-end speedup of \proj over the baseline algorithms on mobile Ampere GPU Orin. \mode{Org.+S} is the baseline that applies sparse pixel sampling without our pixel-based rendering pipeline.
    Note that, the tracking speedup is aligned with the end-to-end speedup.
    }
    \label{fig:gpu_e2e_perf}
\end{figure}

We first show that our sparse pixel sampling with proposed pixel-based rendering can already achieve significant speedups and energy savings on off-the-shelf GPUs \textit{without} hardware support.
\Fig{fig:gpu_e2e_perf} shows the end-to-end speedup and energy savings of \proj on the Nvidia Orin SoC.

In our evaluation, we assume that tracking and mapping are executed separately on two identical mobile GPUs with equal compute power, allowing these two stages to run in parallel.
For comparison, we also include a variant, \mode{Org.+S}, which applies only our sparse pixel sampling algorithm without integrating our pixel-based rendering pipeline.

\paragraph{End-to-End Performance.}
\Fig{fig:gpu_e2e_perf} shows the end-to-end speedup and normalized energy of \proj compared to baseline algorithms on a mobile Ampere GPU on Nvidia Orin.
The left y-axis shows the speedup against the GPU baselines, while the right y-axis shows the normalized energy.
Overall, \proj achieves 14.6$\times$ speedup and saves 86.1\% energy compared to the original algorithms.
The end-to-end speedup is aligned with the tracking speedup, as the latency of mapping can be hidden by tracking.
In comparison, \mode{Org.+S} achieves only 3.4$\times$ speedup and 55.5\% energy savings. 
This is because the original tile-based rendering pipeline would result in low GPU utilization and severe warp divergence.

% \begin{figure}[t]
% \centering
% \begin{minipage}[t]{0.48\columnwidth}
%   \centering
%   \includegraphics[width=\columnwidth]{gpu_tracking_perf}
%   \caption{The speedup and energy savings comparison on tracking.}
%   \label{fig:gpu_tracking}
% \end{minipage}
% \hspace{1pt}
% \begin{minipage}[t]{0.48\columnwidth}
%   \centering
%   \includegraphics[width=\columnwidth]{gpu_mapping_perf}
%   \caption{The speedup and energy savings comparison on mapping.}
%   \label{fig:gpu_mapping}
% \end{minipage}
% \end{figure}

% \begin{figure}[t]
%     \centering
%     \includegraphics[width=\columnwidth]{gpu_tracking_mapping_perf}
%     \caption{The speedup and energy savings comparison on tracking and mapping. Tracking often achieves much higher speedup and energy savings due to sparser pixel sampling. Numbers are normalized against GPU baselines.}
%     \label{fig:gpu_tracking_mapping}
% \end{figure}

\begin{figure}[t]
\centering
\begin{minipage}[t]{0.48\columnwidth}
  \centering
  \includegraphics[width=\columnwidth]{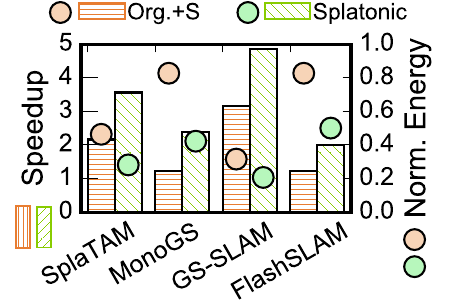}
  \caption{The speedup and energy savings comparison on mapping. Despite limited speedup, the latency of mapping still can be hidden by tracking.}
  \label{fig:gpu_mapping}
\end{minipage}
\hspace{1pt}
\begin{minipage}[t]{0.48\columnwidth}
  \centering
  \includegraphics[width=\columnwidth]{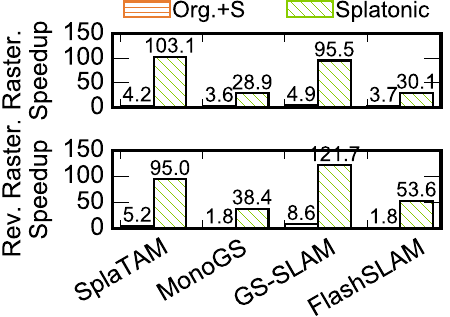}
  \caption{The speedup on two key bottleneck stages, rasterization in forward pass and reverse rasterization in the backward pass, during tracking.}
  \label{fig:gpu_bottleneck_speedup}
\end{minipage}
\end{figure}

Meanwhile, we also report the standalone speedup and energy savings for mapping in \Fig{fig:gpu_mapping}.
On average, \proj can achieve only 3.2$\times$ speedup and 60.0\% energy savings on mapping. 
This is because mapping needs to render more pixels (roughly one pixel per $4\times4$ tile) to reconstruct unseen regions.
As the number of rendered pixels increases, the advantages of our pixel-based pipeline might be offset by the additional overhead introduced in projection and sorting stages due to no sharing of computation between pixels.
\Sect{sec:eval:sens} further shows the sensitivity of speedup to the pixel sampling rate.

% \begin{figure}[t]
%     \centering
%     \includegraphics[width=\columnwidth]{gpu_bottleneck_speedup}
%     \caption{The speedup on two key bottleneck stages, rasterization in forward pass and reverse rasterization in the backward pass, during tracking.
%     % Numbers are normalized against the GPU baselines.
%     }
%     \label{fig:gpu_bottleneck_speedup}
% \end{figure}

\paragraph{Bottleneck stages.}
The main performance gain of \proj on GPU is from addressing two key bottlenecks, rasterization in forward pass and reverse rasterization in the backward pass.
\Fig{fig:gpu_bottleneck_speedup} further analyzes the speedup of these two bottleneck stages during the tracking process.
Without modifying the pipeline, applying sparse sampling alone yields only 4.1$\times$ and 4.3$\times$ speedup on rasterization and reverse rasterization, respectively.
% Because, in the conventional pipeline, only one thread in a warp performs meaningful work while the rest remain idle.
In contrast, our pipeline achieves 64.4$\times$ and 77.2$\times$ speedup on these two stages, respectively.
% Because our pipeline eliminates the warp divergence and maximizes the GPU utilization.

\begin{figure}[t]
    \centering
    \includegraphics[width=\columnwidth]{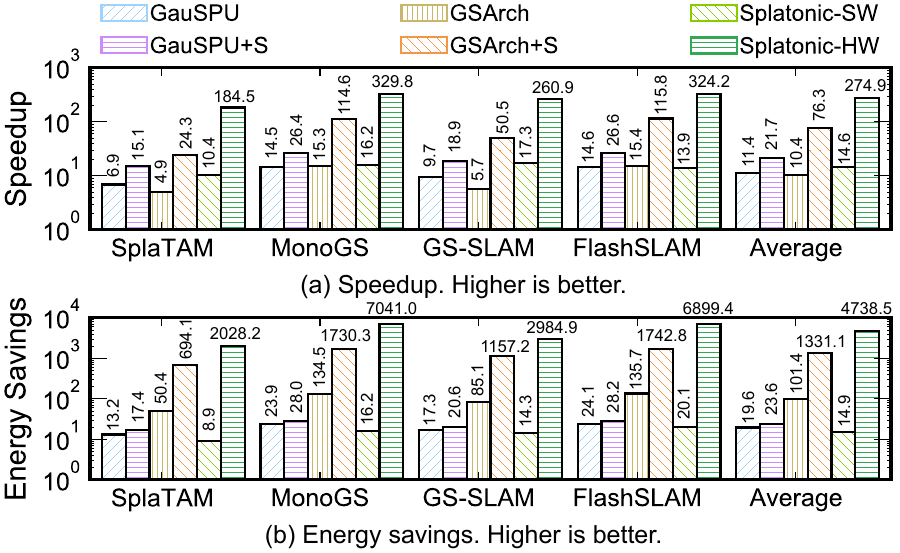}
    \caption{The performance and energy consumption comparison across different dedicated architectures during tracking.
    ``S'': applying sparse pixel sampling.
    Numbers are normalized against \mode{GPU}.
    }
    \label{fig:hw_tracking_perf}
\end{figure}

\subsection{Hardware Performance}
\label{sec:eval:hw}

Since tracking dominates the overall execution, we primarily focus on the tracking performance in this section.
\Fig{fig:hw_tracking_perf} shows the comparison of the performance and energy savings across different architectures.
For a fair comparison, we also include variants of \mode{GauSPU} and \mode{GSArch} that incorporate our sparse sampling algorithm, denoted with the ``+S'' suffix.
All numbers are normalized to the GPU baseline on Orin.
To better show the difference across hardware variants, we define the energy saving as the ratio of energy consumption between \mode{GPU} and each corresponding variant.

In \Fig{fig:hw_tracking_perf}a, \mode{\proj-hw} achieves 274.9$\times$ speedup, the highest performance compared to \mode{GauSPU+S} and \mode{GSArch+S}.
Because both \mode{GauSPU+S} and \mode{GSArch+S} are accelerators designed for tile- or subtile-based rendering, where sparse pixel sampling leads to poor PE utilization in rasterization and reverse rasterization stages.
Also, our simplified rasterization engine further reduces both the overall computation and off-chip data traffic in these two stages.
Surprisingly, our software version, \mode{\proj-sw}, outperforms \mode{GauSPU} and \mode{GSArch}, both of which are variants that execute the dense 3DGS-SLAM.

On energy savings, \Fig{fig:hw_tracking_perf}b shows a similar trend with the performance results.
Overall, \mode{\proj-hw} has the highest energy efficiency with 4738.5$\times$ of energy savings.
In comparison, \mode{GauSPU+S} and \mode{GSArch+S} achieves 23.6$\times$ and 1331.1$\times$ of energy savings, respectively.
The relatively low energy efficiency of \mode{GauSPU+S} is because it relies on GPUs to execute the projection and sorting stages. 
Meanwhile, \mode{GSArch+S} has relatively smaller energy savings compared to \mode{\proj-hw} due to its sub-tile rendering, whereas \mode{\proj-hw} benefits from a simplified rasterization.

\begin{figure}[t]
\centering
\begin{minipage}[t]{0.48\columnwidth}
  \centering
  \includegraphics[width=\columnwidth]{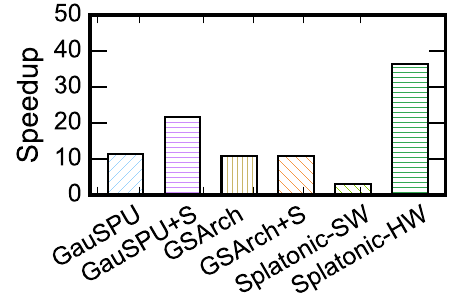}
  \caption{The mapping speedup comparison across different dedicated architectures. \proj still outperforms the other two accelerators. The legend is shared with \Fig{fig:hw_tracking_perf}.}
  \label{fig:hw_mapping_perf}
\end{minipage}
\hspace{1pt}
\begin{minipage}[t]{0.48\columnwidth}
  \centering
  \includegraphics[width=\columnwidth]{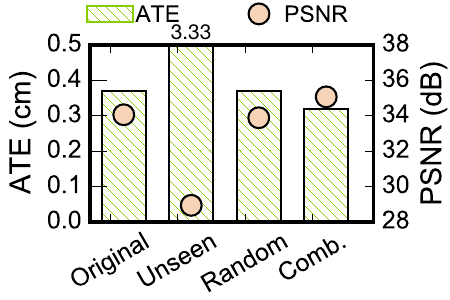}
  \caption{Ablation study of different sampling strategies in mapping. 
  We only show the results of SplaTAM. 
  ``Comb'': uses both weighted random sampling and unseen pixels.}
  \label{fig:ablation}
\end{minipage}
\end{figure}

In addition, we also show the mapping performance comparison in \Fig{fig:hw_mapping_perf}.
The overall trends stay the same as tracking.

\subsection{Ablation Study}
\label{sec:eval:abl}

\Fig{fig:ablation} shows an ablation study that evaluates the contributions of different components in our mapping sampling algorithm. 
The results show that combining weighted random sampling (for texture-rich pixels) with unseen pixels yields the highest accuracy. 
Note that, our combined strategy outperforms the original algorithm in both pose tracking and reconstruction quality. 
Overall, our combined variant achieves a 0.05~cm reduction in pose error and a 1.0~dB quality improvement compared to the baseline algorithm.

\begin{figure}[t]
\centering
\begin{minipage}[t]{0.48\columnwidth}
  \centering
  \includegraphics[width=0.9\columnwidth]{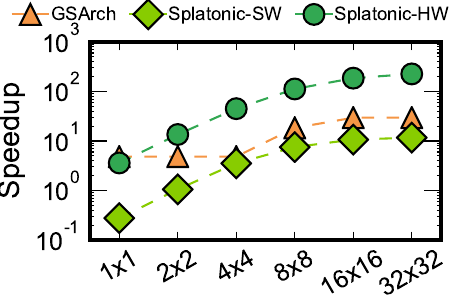}
  \caption{The sensitivity of performance to the sampling rate. Numbers are normalized to \mode{GPU}.}
  \label{fig:sens_speedup}
\end{minipage}
\hspace{1pt}
\begin{minipage}[t]{0.48\columnwidth}
  \centering
  \includegraphics[width=0.9\columnwidth]{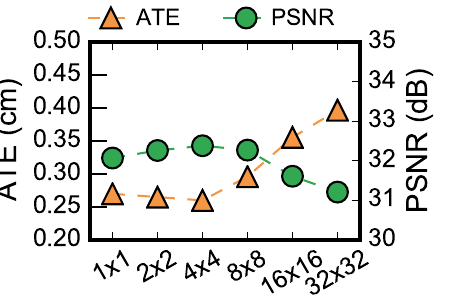}
  \caption{The sensitivity of accuracy to the sampling rate.}
  \label{fig:sens_acc}
\end{minipage}
\end{figure}

\begin{figure}[t]
\centering
  \includegraphics[width=0.48\columnwidth]{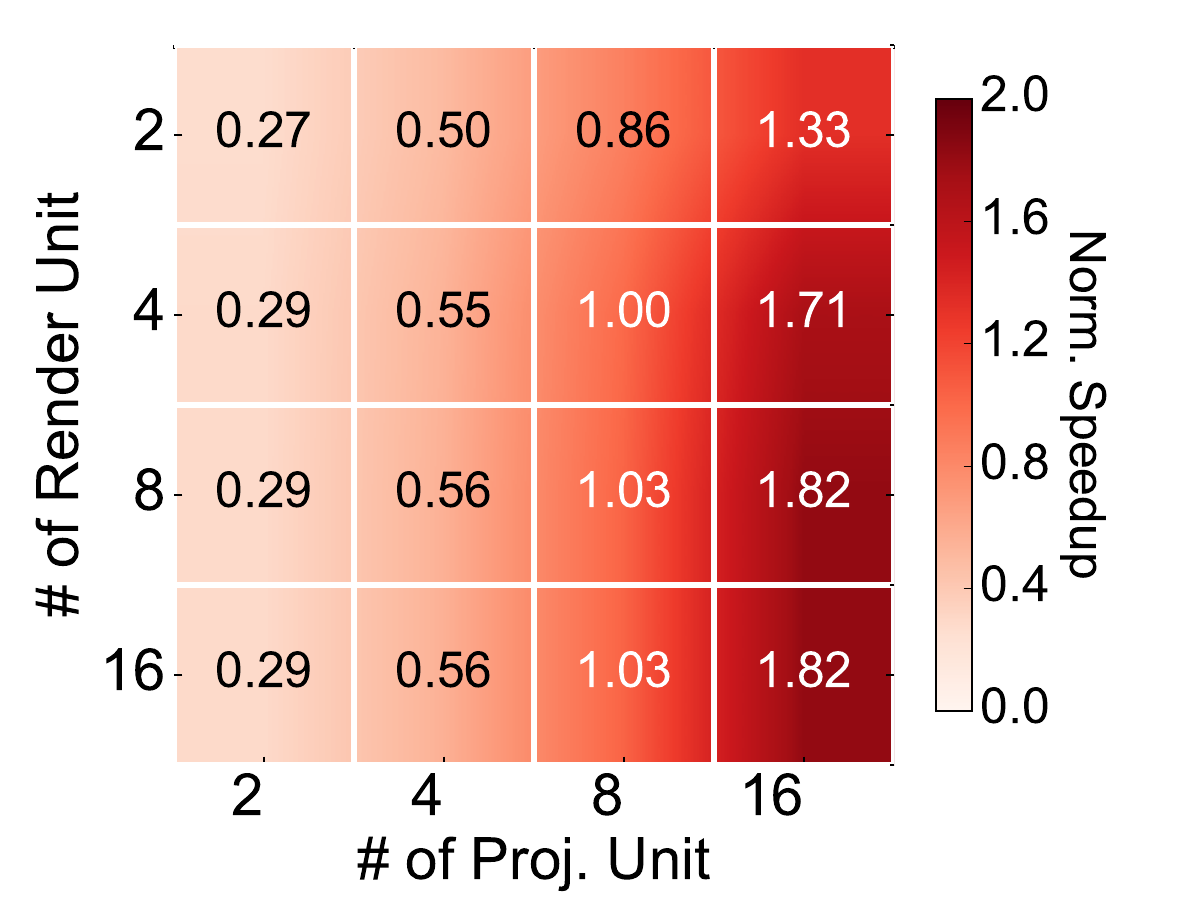}
  \caption{The sensitivity of performance to the number of projection units and render units.}
  \label{fig:sens_pe}
\end{figure}

\subsection{Sensitivity Study}
\label{sec:eval:sens}

\Fig{fig:sens_speedup} shows the sensitivity of performance to the sampling rate. 
The x-axis represents different tile sizes, where each tile contains one sampled pixel. 
The results show that our pixel-based rendering is not always the optimal choice. 
As the tile size decreases, the data sharing among adjacent pixels increases. 
Tile-based rendering can amortize computation, thereby achieving higher speedup. 
For example, at a tile size of $1 \times 1$, \mode{\proj-hw} yields lower speedup than \mode{GSArch}. 
However, when rendered pixels are sparse, \mode{\proj-hw} significantly outperforms tile-based accelerators. 

Meanwhile, \Fig{fig:sens_acc} shows the sensitivity of mapping to the sampling rate. 
Here, we present the SplaTAM result of a single sequence, \textit{Office 2}, in the Replica dataset~\cite{straub2019replica}. 
We find that a tile size of $4 \times 4$ gives the best trade-off between performance and reconstruction accuracy. 
The sensitivity of tracking to the sampling rate has already been shown in \Fig{fig:tracking_sampling}.

% \begin{figure}[t]
% \centering
% \begin{minipage}[t]{0.48\columnwidth}
%   \centering
%   \includegraphics[width=\columnwidth]{sens_pe}
%   \caption{\hl{The sensitivity of performance to the number of projection unit and render unit.}}
%   \label{fig:sens_pe}
% \end{minipage}
% \hspace{1pt}
% \begin{minipage}[t]{0.48\columnwidth}
%   \centering
%   \includegraphics[width=\columnwidth]{high_freq_speedup}
%   \caption{\hl{The performance comparison at 1GHz clock frequency.}}
%   \label{fig:high_freq}
% \end{minipage}
% \end{figure}

{\Fig{fig:sens_pe}} shows the sensitivity of performance to the number of projection units and render units.
We do not conduct the sensitivity study on different buffer sizes, as each buffer size is tightly coupled with the corresponding PEs to support double buffering without wasting the on-chip buffer resources.
In {\Fig{fig:sens_pe}}, the buffer size of each configuration is proportional to its corresponding PE counts. 
All performance numbers are normalized to the default configurations, 8 projection units and 4 render units.
Overall, we find that the performance gain is mainly affected by the number of projection units, especially when there is a small number of projection units.
As the number of projection units increases, projection is no longer a bottleneck.
Instead, further increasing the number of render units improves the overall performance.

% \subsection{Performance at High Clock Frequency}
% \label{sec:eval:clock}

% \hl{
% {\Fig{fig:high_freq}} shows the performance comparison of \mbox{\proj} when clocked at a higher frequency of 1~GHz.
% Overall, the performance doubles as the frequency increases from 500~MHz to 1~GHz.
% This linear scaling is attributed to the compute-intensive nature of the 3DGS pipeline.
% As a result, increasing the clock frequency yields a near-linear performance gain.

% }

\section{Related Work}
\label{sec:related}

\paragraph{3DGS Acceleration.}
There is a wide range of studies that have proposed different techniques to improve the efficiency of 3DGS on GPUs.
Several studies have proposed various compression schemes or data structures to optimize storage and performance\mbox{~\cite{girish2024eagles, lee2024compact, niemeyer2024radsplat, ren2024octree, kerbl2024hierarchical, liu2025voyager}}.
For instance, CompactGS\mbox{~\cite{lee2024compact}} proposes to use vector quantization to reduce the model size.
While HierarchicalGS\mbox{~\cite{kerbl2024hierarchical}} and OctreeGS\mbox{~\cite{ren2024octree}} leverage the tree-like data structure to avoid unnecessary computations.
Other approaches propose various pruning techniques\mbox{~\cite{fan2023lightgaussian, fang2024mini, mallick2024taming, lin2025metasapiens}} to eliminate insignificant Gaussians.
Additional efforts\mbox{~\cite{feng2024flashgs, gui2024balanced, huang2025seele, wang2024adr}} focus on GPU-level optimizations to mitigate workload imbalance and warp divergence during rasterization. 
For instance, both AdR-Gaussian\mbox{~\cite{wang2024adr}} and Seele\mbox{~\cite{huang2025seele}} tame the warp divergence with co-training techniques.

In contrast to those application-agnostic methods, \proj exploits the unique algorithmic characteristics of SLAM by introducing a pixel sampling algorithm and a pixel-based rendering pipeline, significantly improving efficiency.

\paragraph{Architecture for Neural Rendering.}
A wide range of studies have been done on accelerating neural rendering.
Earlier work primarily focused on Neural Radiance Fields (NeRF)~\cite{lee2023neurex, rao2022icarus, li2023instant, mubarik2023hardware, li2022rt, fu2023gen, feng2024cicero, song2024srender, liu2025cambricon}, which is the predecessor of 3DGS.
Recent years, due to the superior efficiency of 3DGS, studies have shifted their focus to 3DGS accelerations~\cite{feng2025lumina, ye2025gaussian, lee2024gscore, li2025uni, lee2025vr, lin2025metasapiens, durvasula2025arc, he2025gsarch, li2023sltarch, zhang2025streaming}.

Some studies propose dedicated hardware accelerators.
For instance, both GSCore~\cite{lee2024gscore} and GBU~\cite{ye2025gaussian} are designed for accelerating the forward pass in 3DGS.
MetaSapiens~\cite{lin2025metasapiens} leverages the human visual perception and designs a dedicated rendering system for VR. 
Lumina~\cite{feng2025lumina} proposes a caching technique to amortize the per-pixel rendering cost.
Meanwhile, GSArch\mbox{~\cite{he2025gsarch}} and GauSPU\mbox{~\cite{wu2024gauspu}} focus on the training procedure and address the frequent pipeline stalls due to off-chip traffic.
On the other side of the spectrum, some approaches augment general-purpose GPUs to improve 3DGS performance. 
For example, VR-Pipe~\cite{lee2025vr} improves GPU-based inference, while ARC~\cite{durvasula2025arc} proposes architectural optimizations for training.

Nevertheless, all these architectural designs are still built on top of the conventional tile-based rendering paradigm, which is inherently inefficient for sparse pixel processing. In contrast, our co-designed pipelined architecture eliminates redundant computation and significantly improves PE utilization.
Meanwhile, \mbox{\proj} can extend beyond SLAM applications.
Recent studies have explored sparse training techniques~\mbox{\cite{chen2025dashgaussian, ren2024octree}} to reduce the computational overhead of 3DGS training.
Other studies have proposed sparse rendering approaches for foveated rendering in VR~\mbox{\cite{lin2025metasapiens, deng2022fov, feng2024potamoi}}.
By leveraging our pixel-based rendering pipeline, these methods can be further accelerated, demonstrating the broader applicability of our design across diverse 3DGS-related domains.

\paragraph{Sparse Processing.}
Quite a few studies have proposed sparse processing techniques in various vision tasks to reduce the overall computation and data transmission~\cite{kong2016hypernet, ren2015faster, girshick2015fast, girshick2014rich, he2017mask, kodukula2021rhythmic, feng2022real, feng2024blisscam}.
For instance, fast R-CNN~\cite{girshick2015fast} and faster R-CNN~\cite{ren2015faster} introduced region proposal networks for general object detection.
Others~\cite{kodukula2021rhythmic, mudassar2019camera, feng2022real} react based on the previous results or intermediate data to pre-filter related data before sending it through compute-heavy backbones.
For instance, Rhythmic pixel regions~\cite{kodukula2021rhythmic} provides a flexible interface that allows applications to dynamically select region-of-interests.
More recent efforts~\cite{young2019data, feng2023learned, feng2024blisscam, ma2022hogeye} also co-design algorithms with the camera sensor to reduce the overall energy consumption across the entire sensor-compute system.

\section{Conclusion and Discussion}
\label{sec:conc}

In this work, we present \proj, a hardware-software co-designed solution to accelerate 3DGS-based SLAM for real-time performance on mobile platforms. 
By leveraging classic algorithmic insights from traditional SLAM with our dedicated pixel-based rendering, we show that \proj achieves one order of magnitude higher performance on off-the-shelf GPUs, and we further boost performance and efficiency with our dedicated hardware support.
\section*{Acknowledgment}

We are sincerely grateful to Weikai Lin from University of Rochester for providing valuable support throughout this research.
This work was supported by the National Natural Science Foundation of China (NSFC) Grants (62532006 and 62402312) and Shanghai Qi Zhi Institute Innovation Program SQZ202316.

% \clearpage
%%%%%%% -- PAPER CONTENT ENDS -- %%%%%%%%

%%%%%%%%% -- BIB STYLE AND FILE -- %%%%%%%%
\bibliographystyle{IEEEtranS}
\bibliography{refs}
%%%%%%%%%%%%%%%%%%%%%%%%%%%%%%%%%%%%

\end{document}